\documentclass[useAMS,usenatbib]{mn2e}
\usepackage{graphicx, color}
\usepackage{amssymb, natbib, bm}
\title[Gravitational collapse and structure of DM haloes]{What sets the central structure of dark matter haloes?}
\author[Ogiya and Hahn]{Go Ogiya$^{1}$\thanks{E-mail: Go.Ogiya@oca.eu} and Oliver Hahn$^{1}$\thanks{E-mail: Oliver.Hahn@oca.eu}\\
$^{1}$Laboratoire Lagrange, Universit\'e C\^ote d'Azur, Observatoire de la C\^ote d'Azur, CNRS, \\ \ \ \ Blvd de l'Observatoire, CS 34229, F-06304 Nice cedex 4, France
}
\begin{document}

\date{Accepted ***. Received ***; in original form 2017 ***}
\pagerange{\pageref{firstpage}--\pageref{lastpage}} \pubyear{2017}
\maketitle

\label{firstpage}
\begin{abstract}
  Dark matter (DM) haloes forming near the thermal cut-off scale of the density perturbations are unique, since they are the smallest objects and form through monolithic gravitational collapse, while larger haloes contrastingly have experienced mergers. While standard cold dark matter (CDM) simulations readily produce haloes that follow the universal Navarro-Frenk-White (NFW) density profile with an inner slope, $\rho \propto r^{-\alpha}$, with $\alpha=1$, recent simulations have found that when the free-streaming cut-off expected for the CDM model is resolved, the resulting haloes follow nearly power-law density profiles of $\alpha\sim1.5$. In this paper, we study the formation of density cusps in haloes using idealized $N$-body simulations of the collapse of proto-haloes. When the proto-halo profile is initially cored due to particle free-streaming at high redshift, we universally find $\sim r^{-1.5}$ profiles irrespective of the proto-halo profile slope outside the core and large-scale non-spherical perturbations. Quite in contrast, when the proto-halo has a power-law profile, then we obtain profiles compatible with the NFW shape when the density slope of the proto-halo patch is shallower than a critical value, $\alpha_{\rm ini} \sim 0.3$, while the final slope can be steeper for $\alpha_{\rm ini}\ga 0.3$. We further demonstrate that the $r^{-1.5}$ profiles are sensitive to small scale noise, which gradually drives them towards an inner slope of $-1$, where they become resilient to such perturbations. We demonstrate that the $r^{-1.5}$ solutions are in hydrostatic equilibrium, largely consistent with a simple analytic model, and provide arguments that angular momentum appears to determine the inner slope. 
\end{abstract}


\section{Introduction}
\label{sec:intro}
In the standard model of cosmic structure formation, the formation and evolution of structures in the Universe is driven by the dynamics of dark matter (DM), most commonly assumed to be in the form of collisionless cold dark matter (CDM). Gravitationally collapsed and virialised DM haloes drive the formation and evolution of galaxies and galaxy clusters by attracting baryonic gas into their centres where stars form from the condensed gas \citep[e.g.][]{1978MNRAS.183..341W}. Due to the shape of the spectrum of density fluctuations, structures grow hierarchically from the first generation of smallest DM haloes to galaxy clusters, and even larger super-structures today. This theoretical scenario has had tremendous success in reproducing a wealth of observational results, and particularly so on large scales \citep[$\ga$ 1Mpc; e.g.][]{2004ApJ...606..702T}. At the same time, the particle nature of DM still eludes us. DM that is mainly in the form of the lightest stable supersymmetric particle  at the mass scale of several hundred GeV has been a favourite candidate (the weakly interacting massive particle, WIMP). This was mainly motivated by the WIMP miracle where weak-scale interactions would predict the required relic abundance \citep[see e.g.][and references therein]{1996PhR...267..195J}. However, with DM searches still having found no trace of the WIMP, and the well known possible small-scale problems of CDM \citep[see e.g.][and references therein]{2015PNAS..11212249W, 2017arXiv170704256B}, alternative models, such as e.g. warm dark matter (WDM), are also of great current interest.

The non-linear evolution of DM as a collisionless self-gravitating fluid is governed by the Vlasov-Poisson (VP) system of equations. The numerical evolution of the cosmological VP equations is most commonly performed through $N$-body simulations. Thanks to the rapid evolution of both computational facilities and simulation techniques in the last four decades, cosmological $N$-body simulations have been the most powerful tools to model the non-linear phase of structure formation in the Universe \citep[e.g.][]{1975PASJ...27..333M, 1979MNRAS.187..117E, 1979ApJ...228..664A,1983MNRAS.204..891K, 1985ApJ...292..371D, 1991ApJ...370L..15S, 1991ApJ...378..496D, 2005Natur.435..629S}. One of the key results of such simulations has been the discovery of the universal form of the spherically averaged radial density profile of relaxed DM haloes.
\citet[][hereafter NFW]{1996ApJ...462..563N, 1997ApJ...490..493N} found that independently of halo mass, the density profile is well described by the two-parameter family of NFW profiles, given by
\begin{equation}
\rho(r) = \frac{\rho_{\rm s}}{(r / r_{\rm s}) [1 + (r / r_{\rm s})]^{2}}, \label{eq:nfw} 
\end{equation}
where $r$ is the distance from the centre of the DM halo, and $\rho_{\rm s}$ and $r_{\rm s}$ represent the scale density and length, respectively. 
While subsequent studies pointed out some deviations from universality \citep[e.g.][]{1997ApJ...477L...9F, 1999MNRAS.310.1147M, 2000ApJ...529L..69J, 2003MNRAS.344.1237R, 2004ApJ...607..125T}, the NFW profile works remarkably well over a wide range of halo mass, from galactic to cluster scales even if the slope of the power spectrum is changed \citep[e.g.][]{1996MNRAS.281..716C,1997MNRAS.286..865T,2017MNRAS.465L..84L}. 
More recent papers based on higher resolution simulations \citep[e.g.][]{2004MNRAS.349.1039N, 2008MNRAS.391.1685S, 2009MNRAS.398L..21S, 2010MNRAS.402...21N, 2014MNRAS.441.3359D} found that the Einasto profile \citep{1965TrAlm...5...87E} provides however a better fit for haloes in their simulations, especially in the very central region of haloes, $r \la 0.01 r_{\rm vir}$, where $r_{\rm vir}$ represents the virial radius of the DM halo.

Simulations that take into account the damping of perturbations on small scales due to the free streaming of DM particles associated with their kinetic temperature have yielded a less coherent picture. The small but finite kinetic temperature results in a cut-off in the matter power spectrum compared to the perfectly cold limit. This cut-off directly sets the mass scale of the smallest DM haloes by suppressing structure formation below this scale. We note that due to the huge number of particles that would be needed to sample a true warm phase space distribution, modelling DM with a finite temperature in the cold limit while adding the cut-off to the perturbation spectrum is the common and reasonable approach.
For a CDM particle with mass of order $\sim100$~GeV \citep[e.g.][]{2002PhRvD..66a0001H, 2003PhLB..565..176E}, the corresponding lower mass limit of CDM haloes is about an earth mass ($M_{\rm \oplus} \sim 10^{-6} M_{\rm \odot}$). These are the so-called microhaloes (\citealp[e.g.][]{2004MNRAS.353L..23G}, but see also e.g. \citealt{2009NJPh...11j5027B}). 
Microhaloes are expected to be very abundant in galactic- and cluster sized larger DM haloes and may be a significant source of $\gamma$-rays from annihilating DM (\citealp[e.g.][]{2003PhRvD..68j3003B, 2005Natur.433..389D}). 

Thermally produced DM particles with smaller masses have a higher kinetic temperature and thus suppress the growth of fluctuations to larger scales. The resulting truncation of the matter power spectrum hence corresponds to a higher mass of the first dark haloes that are able to form. A particularly interesting regime is of course the WDM regime in which this mass scale coincides with the scale of dwarf galaxies, $\sim 10^{8-9} M_{\rm \odot}$ \citep{2001ApJ...556...93B}. Because of the lack of substructures smaller than this scale, WDM is a possible solution for the so-called missing satellite problem \citep{1999ApJ...522...82K, 1999ApJ...524L..19M}. From recent observational constraints \citep[e.g.][]{2013PhRvD..88d3502V, 2016ApJ...825L...1M}, an acceptable mass of a WDM particle is $\gtrsim 3 {\rm keV}$, which still results in a moderate suppression of the abundance of dwarf galaxies. 

The spherically averaged density profile of DM haloes in models with a resolved cut-off in the matter power spectrum has also been investigated using cosmological $N$-body simulations. Quite in contrast to the perfectly cold simulations, the results have been less conclusive and therefore more controversial. On the one hand, some studies found that the density structure of DM haloes in cosmological $N$-body simulations with a cut-off spectrum is well described by the NFW profile: 
e.g. in the case of hot dark matter (HDM), in which the DM candidate is a neutrino with a mass of $\la$ 0.3 eV \citep{2014PhRvL.112e1303B} and the truncation in the matter power spectrum arises on cluster scales \citep[e.g.][]{2009MNRAS.396..709W}; but also in the WDM case \citep[e.g.][]{2001ApJ...559..516A, 2007ApJ...665....1B, 2014MNRAS.439..300L}. 
On the other hand, \cite{2010ApJ...723L.195I} and \cite{2013JCAP...04..009A} found that microhaloes have steeper central cusps than those of NFW haloes, i.e. follow over a large range of radii more closely a single power-law profile  of the form
\begin{equation}
\rho \propto r^{-\alpha} \ \ \ (\textrm{with }\alpha > 1). 
\end{equation} 

Subsequent studies \citep{2014ApJ...788...27I, 2016arXiv160403131A} confirmed the existence of cusps with $\alpha > 1$ and also demonstrated that the central density slope of the smallest microhaloes is described by $\alpha = 1.5$ for some time after their formation and then becomes increasingly shallower as their masses grow. 
In contrast to previous studies that consistently found NFW profiles in WDM simulations, \cite{2015MNRAS.450.2172P} found a steeper inner profile, but the difference from the NFW profile seems to be less obvious than those seen in microhaloes.

\cite{2016MNRAS.461.3385O} and \cite{2016arXiv160403131A} argued that violent relaxation \citep{1967MNRAS.136..101L} driven by major mergers between microhaloes -- expected to occur frequently because of the shallow slope of the power spectrum above the cut-off and the lack of smaller substructures -- plays a role in making the central density profile shallower. They also found that the dynamical impact of consecutive mergers weakens as the logarithmic central density slope, $\alpha$, approaches unity, i.e., the NFW profile, and the density structure becomes more resilient. This might explain why the NFW profile is universally observed in most cosmological simulations. In light of these results, the controversy of halo density profiles in cosmological $N$-body simulations with the cut-off might be related to the WDM and HDM haloes having already experienced mergers such that their central cusp has been already transformed into that of the NFW model. 

There is however another possible effect that is harder to quantify. It has been known for a long time that $N$-body simulations, in particular those that are initialised from a truncated density spectrum where the cut-off is (well) resolved, suffer from so-called artificial fragmentation \citep[e.g.][]{1997ApJ...479L..79M,1998ApJ...497...38S,2007MNRAS.380...93W,2013MNRAS.434.1171H,2013MNRAS.434.3337A, 2014MNRAS.439..300L}: while the formation of DM haloes should be strongly suppressed below the cut-off scale, spurious haloes appear abundantly along filaments in such simulations. As they accrete onto the first generation of true haloes, they may alter the density structure of physical DM haloes by heating up their central part through mergers and/or by introducing additional noise. 

Since the first generation of haloes are the seeds (or building blocks) of all larger systems formed later, such as galaxies and galaxy clusters, in the hierarchical structure formation scenarios, investigating the origin of their structure and untangling physical from numerical effects is of key importance to understand the processes of formation and evolution of cosmic structures.
Especially in the WDM model, a part of them would be expected to survive to late times as the surroundings of dwarf galaxies which are dynamically dominated by DM and might retain the memory of their formation processes.

The role of the cut-off in the spectrum as well as small scale perturbations in setting the density profile of DM haloes is therefore still unclear. With the profiles in at least some cosmological scenarios {\em not} following a universal NFW profile, we are thus led to ponder the following key questions: 
\begin{enumerate}
\item Why do CDM haloes in cosmological simulations with the cut-off in the matter power spectrum, i.e. microhaloes, have steeper density cusps ($\alpha > 1$) than those in cosmological simulations without the cut-off ($\alpha = 1$, i.e., the NFW profile)? 
\item Why are cusps with $\alpha > 1$ seen in microhaloes more obviously than in WDM or HDM haloes? 
\item And thus ultimately: What is the role of the peak profile and shape of the proto-halo patch at variance with the presence of small scale perturbations in setting the inner profile?
\end{enumerate}

It is clear that DM haloes near the cut-off in the matter power spectrum form differently from their equal-mass counterparts in perfectly cold DM: rather than by a sequence of hierarchical mergers, they form through monolithic gravitational collapse, and such DM haloes of the first generation have not experienced any mergers. Their collapse originates from an asymmetric perturbation of an otherwise smooth density field. The main density peak of a first-generation halo forming at some mass scale $M$ is smooth below that scale which leads to two (interrelated) main differences with respect to the peak properties in the cold limit: (1) when smoothed on a scale much smaller than $M$, the density at the centre of the peak does not increase, and (2) there are no small-scale density fluctuations. Based on these two aspects, we will study the differences in the resulting density profiles after the collapse of idealised proto-halo peaks in this paper and investigate the impact of the peak profile, non-spherical perturbations and small-scale noise on the outcome. This will allow us to explore the origin of the very steep cusps observed in cosmological simulations using idealised $N$-body simulations with high resolution.

Our paper is structured as follows.  In Section~\ref{sec:model}, we give an overview over previous analytical results and explain the model that we adopt in our subsequent simulations. 
Then, in Section~\ref{sec:results}, we present the results of our suite of numerical simulations. 
We discuss some phenomenological arguments to explain the simulation results in Section \ref{sec:phenom_model}, before we conclude in Section~\ref{sec:summary}. 

\section{Model Description}
\label{sec:model}
In this section, we briefly review previous results on the collapse of spherical density perturbations under self-gravity in an expanding universe. Then, we discuss in detail the initial conditions that we will adopt in our subsequent simulations as model systems for the gravitational collapse of a proto-halo patch to a DM halo.
      
\subsection{Spherical infall models}
\label{subsec:infall}
Both analytical studies and numerical simulations have provided important insights into the formation of DM haloes and their internal structure.
Early analytical studies focused on the gravitational collapse of an isolated spherical overdense patch in an Einstein-de~Sitter universe \citep[e.g.][]{1972ApJ...176....1G, 1975ApJ...201..296G}. In such spherical infall models, collisionless material (DM) expands first with the Hubble flow and then falls back towards the centre of the isolated system, thereby establishing a density profile of functionally stable form with a singular central density cusp by undergoing multiple shell-crossing. 

\cite{1984ApJ...281....1F} subsequently developed a more general spherical infall model in which the initial mass perturbation, $M_{\epsilon}$, is described by a parameter, $\epsilon$, as $M_{\epsilon}(r)/M_0(r) \propto M_0^{-\epsilon}(r)$, where $M_0(r)$ is the unperturbed mass enclosed within the radius, $r$. 
The parameter is defined in the range of $0 < \epsilon \leq 1$ and the radial dependence of the perturbation density, $\rho_{\epsilon}(r)$, is given as 
\begin{equation}
\rho_{\epsilon}(r) = \frac{1}{4 \pi r^2} \frac{d M_{\epsilon}(r)}{dr} \propto r^{-3 \epsilon}. \label{eq:epsilon}
\end{equation}
\cite{1984ApJ...281....1F} then derived the self-similar solutions for collapsed systems and the asymptotic behaviour of the central density structure as $\rho(r) \propto r^{-2}$ for $0 < \epsilon \leq 2/3$ (see Appendix \ref{subsec:rad_orb}), and as $\rho(r) \propto r^{-9 \epsilon/(1+ 3 \epsilon)}$ for $2/3 \leq \epsilon \leq 1$. 
\cite{1985ApJS...58...39B} obtained the whole structure of the collapsed systems for the model with $\epsilon = 1$.
He also derived the density structure of $\alpha = 1.5$ for collapsing pressureless fluids, but the solution is valid only before shell crossing.
\cite{1995PhyU...38..687G} obtained a density structure of $\alpha=1.7-1.9$ for proto-halo patches with the initial profile of $\rho_{\rm ini} \propto (1-r^2)$. 
Analyses in these studies were however restricted by the assumption of purely radial orbits. 
Subsequent analytical work also included non-radial orbits in which case the radial dependence of the central density profile is given by $\rho(r) \propto r^{-9 \epsilon/(1+ 3 \epsilon)}$ for $0 < \epsilon \leq 1$ \citep[e.g.][]{1997PhRvD..56.1863S, 2001MNRAS.325.1397N}. 
$N$-body simulations confirmed these analytical predictions \citep{1992ApJ...394....1W, 2006ApJ...653...43M, 2011MNRAS.414.3044V}. 
\cite{2004ApJ...604...18W} suggested that $\alpha \sim 1.5-1.7$ would be the upper limit for the inner density slope of DM haloes in realistic cases.

More recent analytical studies found that the central density structure depends not only on the parameter $\epsilon$ but also on the velocity structure of the gravitationally collapsed systems: those systems that are dynamically dominated by radial velocity components have steeper central density slopes (e.g. \citealp{2000ApJ...538..517S}, \citealp{2006MNRAS.368.1931L}, hereafter L06; \citealt{2011ApJ...743..127L}). Interestingly, these authors predicted the NFW-like central structures, $\rho \propto r^{-1}$, for the models with $\epsilon \la 0.1$ when the collapsed systems have isotropic velocity dispersion \citep[see also][]{1997ApJ...480...36T}. 

In summary, for $\epsilon \la 0.1$ these analytical models predict an inner slope of $\alpha = 1$ in the limit of vanishing velocity anisotropy and an inner slope of $\alpha = 2$ in the limit of purely radial orbits. {\em We thus formulate the hypothesis that for realistic three-dimensional systems these values provide the limiting cases for a range of proto-halo profiles and thus expect that objects collapsing at the free-streaming scale plausibly occupy an intermediate regime $1 < \alpha < 2$.
}
We will next discuss the setup we choose to study this hypothesis. 

\subsection{The proto-halo peak model in this work}
\label{subsec:our_model}
In this paper, we adopt a model of a proto-halo patch of increasing sophistication: first, we set up a simplified spherically symmetric proto-halo profile motivated by the mean peak profile in a Gaussian random field with a power spectrum both with and without a free-streaming cut-off; then, in further steps, we break spherical symmetry by adding nonspherical large-scale potential perturbations in a controlled way, and finally we also add potential fluctuations on scales smaller than the cut-off scale. 

\subsubsection{The spherically symmetric proto-halo profile}
We thus start by considering an isolated spherical overdense region in a flat universe, $\delta(r) \equiv [\rho_{\rm i}(r) - \rho_{\rm b}]/\rho_{\rm b}$, where $\rho_{\rm i}(r)$ and $\rho_{\rm b}$ are the density structure of the overdense region and the cosmic mean matter density, respectively. Under the assumptions that haloes form at local peaks of the overdensity field (equivalent also to local maxima in velocity convergence in linear theory), the peak theory of \citet[][hereafter BBKS]{1986ApJ...304...15B} applies and allows to calculate the mean spherical peak profile (BBKS, equation~7.10) given a filter kernel that determines the scale of the peak. The natural choice for the filter kernel when studying the behaviour of spectra with a cut-off is of course given by the cut-off function itself. We note that a constrained spherical peak in the formalism of \cite{1991ApJ...380L...5H} is equivalent of course to the BBKS formula. However, here, we will adopt a simpler model with almost identical asymptotic behaviour as the mean BBKS profile. We will discuss the difference between our simplified peak profile and the full BBKS profile subsequently. 
When $\delta(r = 0)$ is large enough, the point $r = 0$ is expected to be a local maximum of the density field, and we can assume that the typical structure of the overdense region is sufficiently well described by the two-point correlation function $\xi(r)$, so that
\begin{equation}
  \delta(r) \propto \xi(r).
  \label{eq:delta_xi}
\end{equation} 
Analytical studies based on the spherical infall models mentioned in Section \ref{subsec:infall} and equation (\ref{eq:delta_xi}) investigated the relation between the slope of the power spectrum and the density slope of collapsed haloes and found steeper density slopes for shallower power spectra \citep[e.g.][]{1985ApJ...297...16H}. Subsequently, \cite{2000MNRAS.311..423L} proposed a modified model and found density profiles of haloes that agree well with the NFW model.

We computed two-point correlation functions for the cosmological models of CDM without the cut-off, of CDM with the cut-off, of WDM, and of HDM using a part of the {\sc music} code \citep{2011MNRAS.415.2101H}. The code uses the fitting formulae of \cite{1998ApJ...496..605E} for the transfer function and of \cite{2004MNRAS.353L..23G} and \cite{2001ApJ...556...93B} for computing the power spectrum damped by free streaming motion of CDM and WDM particles. 
We assume CDM (WDM) particles with a mass of 100 GeV (3.5 keV).
The damping factors of the transfer function proposed by \cite{2001ApJ...556...93B} is also employed in the HDM model assuming that HDM particles have a mass of 30 eV. 
The {\it Wilkinson Microwave Anisotropy Probe} (WMAP) 7-year cosmological parameter set is adopted throughout this paper \citep{2011ApJS..192...18K}. \footnote{The cosmological parameters adopted in this study are the Hubble constant at the present time, $H_0=70.3{\rm kms^{-1} Mpc^{-1}}$; rms of density fluctuations measured in the sphere with a radius of $8 h^{-1} {\rm Mpc}$, $\sigma_8=0.809$; power law index of the primordial power spectrum, $n_{\rm s}=0.96$; fraction of the energy of the universe due to total (dark matter + baryon) mass, $\Omega_{\rm m}=0.27$; and fraction of the energy of the universe due to baryon mass, $\Omega_{\rm b}=0.0451$. We assume a flat universe model throughout the paper and the fraction of the energy of the universe due to the cosmological constant is set to be $\Omega_{\rm \Lambda}=1-\Omega_{\rm m}=0.73$.}

\begin{figure}
  \centering 
   \includegraphics[width=85mm]{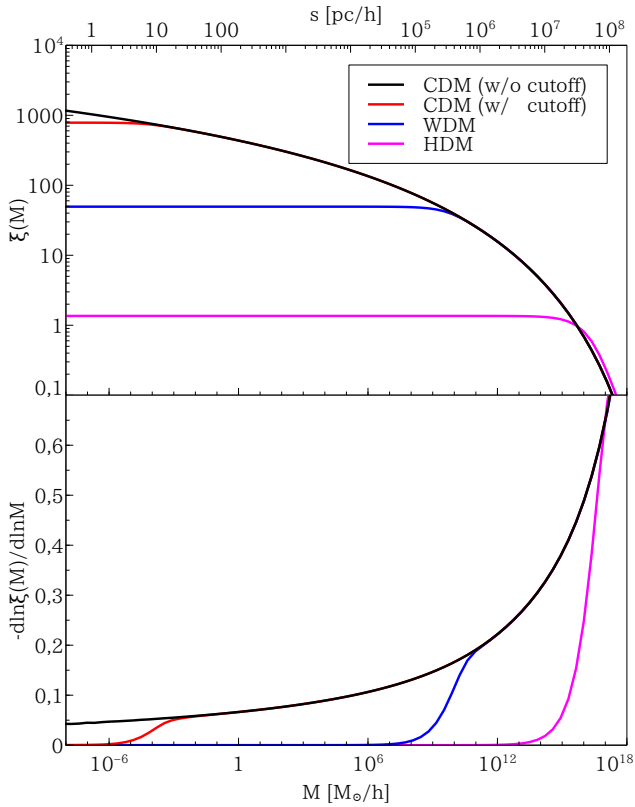}
   \caption{
     {\it Upper panel:} Two-point correlation function as a function of mass scale, $M = (4 \pi/3) \rho_{\rm b, 0} s^3$, where $\rho_{\rm b, 0}$ is the cosmic mean density at the present time.
     The upper horizontal axis represents the corresponding spatial scales, $s$. 
     {\it Lower panel:} Logarithmic slope, -$d \ln{\xi}/d\ln{M}$, which would correspond to the slope of the initial mass perturbation in the spherical infall models. 
       In each panel, black, red, blue and magenta lines represent the models of CDM without the cut-off in the matter power spectrum, of CDM with the cut-off, of WDM and of HDM, respectively. 
       \label{fig:xi}
}
\end{figure}

The upper panel of Figure~\ref{fig:xi} shows the two-point correlation function as a function of mass scale, $M = (4 \pi/3) \rho_{\rm b, 0} s^3$, where $\rho_{\rm b, 0}$ is the cosmic mean density at the present time. 
The corresponding spatial scales, $s$, are shown in the upper horizontal axis. 
Black, red, blue and magenta lines represent the cosmological models of CDM without the cut-off, of CDM with the cut-off, of WDM and of HDM, respectively. 
The two-point correlation function diverges for $r\to 0$ in the model without the cut-off in the matter power spectrum. 
However, in other models in which the cut-off in the matter power spectrum is imposed, the two-point correlation function asymptotes to a constant below the scale of free streaming damping of DM particles since density fluctuations have been smoothed out on these scales. 
To model the core structure in the two-point correlation function, we modify equation (\ref{eq:epsilon}) as 
\begin{equation}
\rho_{\rm \epsilon}(r) \propto (r^2 + r_{\rm c}^2)^{-3 \epsilon/2}, \label{eq:epsilon_rc}
\end{equation}
where $r_{\rm c}$ is the core size. 
This form corresponds to equation~(\ref{eq:epsilon}) in the limit of $r_{\rm c} \to 0$ and ensures the existence of the peak (i.e. vanishing gradient) at the centre, as seen in the two-point correlation functions (coloured lines in Figure \ref{fig:xi}). 
The core size in $\xi(M)$ is approximately the same as the lower mass limit of DM haloes in the model of CDM with cut-off (of WDM), $\sim 10^{-6} (10^8) M_{\rm \odot}$. 
The lower mass limit of DM haloes in the HDM model is $\sim 10^{14}M_{\rm \odot}$.

\begin{table}
\begin{center}
\caption{
Relation between the parameter to control the density slope of the proto-halo patch, $\epsilon$, and redshifts to start and finish the simulations, $z_{\rm i}$ and $z_{\rm f}$. All models follow $(1+z_{\rm i})/(1+z_{\rm f})=12.5$. Assumed DM model, which determines the physical scale of the haloes, is also shown. 
}
\begin{tabular}{cccc}
$\epsilon$ & $z_{\rm i}$ & $z_{\rm f}$ & DM model \\ 
\hline 
0.01             & 399  & 31        & CDM      \\
0.05 (microhalo) & 399  & 31        & CDM      \\
0.17 (WDM)       & 124  &  9        & WDM      \\
0.30             & 124  &  9        & WDM      \\
0.60 (HDM)       & 11.5 &  0        & HDM      \\
0.80             & 11.5 &  0        & HDM      \\
\hline 
\end{tabular}
\label{tab:zi_zf}
\end{center}
\end{table}

The lower panel shows the logarithmic slope of the two-point correlation function. 
The vertical axis in the panel corresponds to the parameter in the spherical infall model, $\epsilon$, when we assume the relation given in equation~(\ref{eq:delta_xi}). 
The typical structure of overdense regions near the free streaming damping scale in the models of CDM with cut-off, of WDM and of HDM can thus be well approximated with the parameters $\epsilon = 0.05$, 0.17 and 0.60, respectively. 
In addition, we also run simulations for $\epsilon = 0.01$, 0.30 and 0.80. Table \ref{tab:zi_zf} describes the redshifts to start and finish the simulations, $z_{\rm i}$ and $z_{\rm f}$. In all simulations, the scale factor of the universe grows by a factor of 12.5 between start and end of the run, i.e. $(1+z_{\rm i})/(1+z_{\rm f})=12.5$.

\begin{figure}
  \centering 
   \includegraphics[width=85mm]{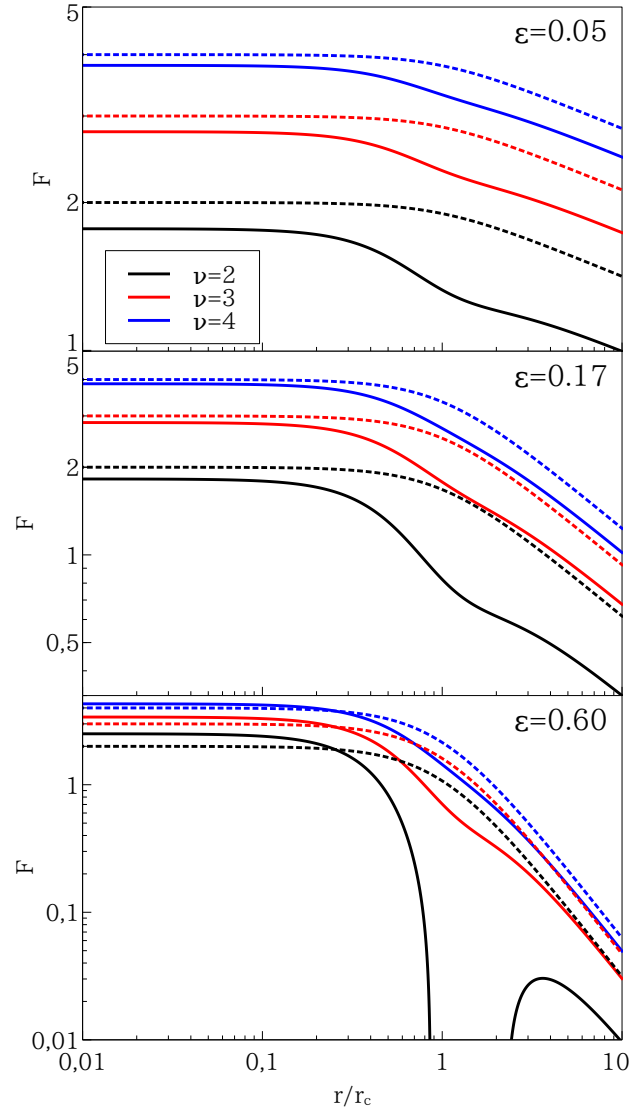}
   \caption{
     Average radial profile of density peaks. 
     The radial bins are scaled by the size of the initial core, $r_{\rm c}$.
     Top, middle and bottom panels show the profiles of peaks described with $\epsilon=0.05, 0.17$ and $0.60$.
     Black, red and blue lines in each panel represent density peaks of the peak height, $\nu=2, 3$ and 4, respectively.
     To draw solid and dotted lines, equations (7.10) and (7.11) in \protect\cite{1986ApJ...304...15B} are used assuming the function form for two-point correlation function, $\xi(r) \propto (r^2+r_{\rm c}^2)^{-3\epsilon/2}$. 
     Solid lines are the predictions by the peak theory of \protect\cite{1986ApJ...304...15B}.
     Our simplified model (dotted lines) reasonably works for density peaks of $\nu \ga 3$. 
     \label{fig:bbks_eq7.10}
    }
\end{figure}

As we have already mentioned above, equation~(\ref{eq:delta_xi}) is only an approximation to the mean peak shape predicted by BBKS. In Figure~\ref{fig:bbks_eq7.10} we show the difference between our simplified peak profile (dotted line) and the corresponding calculation for a mean spherical BBKS peak (solid line). The functional form of equation (\ref{eq:epsilon_rc}), $\xi(r) \propto (r^2+r_{\rm c}^2)^{-3\epsilon/2}$, is assumed for the two-point correlation function to draw the solid lines using equation 7.10 in BBKS. We indeed observe that the density profiles of the two models have similar radial dependencies and the simplified model is clearly reasonable for density peaks of $\nu \ga 3$. Although the core sizes in the two models are slightly different from each other, our results are insensitive to the size of the initial core as shown in Section \ref{subsec:ini_core}. In addition, the simplified model corresponds to equation (\ref{eq:epsilon}) in the limit of $r_{\rm c} \to 0$ studied by previous papers and enables us to do direct comparisons with their results.

Since we assume that $\rho_{\rm \epsilon}$ given in equation~(\ref{eq:epsilon_rc}) is positive, negative terms are also needed to compensate the overdense perturbation at large distances, so that the mean overdensity inside a radius $r$ goes to $0$ as $r\to\infty$. If the perturbation is not compensated, all DM will collapse and belong to the halo after a certain period of time, which is inconsistent with cosmological simulations in which homogeneity is reached by definition at some large scale. We model the necessary negative density perturbation term as
\begin{equation}
\rho_{\rm \gamma}(r) \propto -r^{\gamma}, \label{eq:gamma}
\end{equation}
where $\gamma$ is a parameter, which we fix to $\gamma = 2$ \citep[e.g.][]{1995PhyU...38..687G}. 
Finally, we thus arrive at the full expression for the spherically symmetric proto-halo peak profile, $\rho_{\rm i}(r)$, that we will assume throughout this paper: 
\begin{equation}
\rho_{\rm i}(r) = \rho_{\rm b}(z_{\rm i}) + \rho_{\rm \epsilon}(r) + \rho_{\rm \gamma}(r), \label{eq:ini_tot}
\end{equation}
where $\rho_{\rm b}(z_{\rm i}) = \rho_{\rm b, 0}(1+z_{\rm i})^3$ is the cosmic mean density at the initial redshift of the model, $z = z_{\rm i}$.  

We employ the Zel'dovich approximation \citep{1970A&A.....5...84Z} to perturb a regular particle lattice and thus obtain initial conditions whose density structure follows equation~(\ref{eq:ini_tot}) in the linear regime. 
Adopting a Zel'dovich perturbed regular lattice leads to initial conditions with low enough discreteness noise and a reasonable independence on the initial starting time. We then extract the almost (see also the next subsection) spherical region from the perturbed lattice as the initial particle distribution.
\footnote{We note that the rejection sampling scheme could be considered as another possible way to generate a density perturbation consistent with $\rho_{\rm i}$. However, Poisson noise and the presence of decaying modes are so strong as to render the outcome basically useless.}

The equation of motion of the model naturally includes not only the self-gravity of DM, ${\bf g}$, but also the cosmic expansion, so that
\begin{equation}
\frac{d^2 {\bf r}}{dt^2} = {\bf g} + H_0^2 \Omega_{\rm \Lambda} {\bf r}, 
\end{equation}
where {\bf r} represents the position of a particle in a physical coordinate and $H_0$ and $\Omega_{\rm \Lambda}$ are the Hubble constant and fraction of the energy of the universe due to the cosmological constant at the present time, respectively. We assume a flat universe model so that the fraction of the energy of the universe due to mass is $\Omega_{\rm m} = 1 - \Omega_{\rm \Lambda}$. 
The inclusion of the second term does not significantly alter the density structure of DM haloes.
\footnote{Only the HDM model collapses late enough for $\Omega_{\rm \Lambda}$ to play a role. The only effect we observe is a reduction of the accretion rate, but no effect on the structure of the density profile compared to the other runs.}

Starting from this spherically symmetric model, we allow for two additional sophistication which we will discuss next: non-spherical large-scale symmetry-breaking perturbations, and small-scale Gaussian noise.

\subsubsection{Breaking the spherical symmetry}
\label{sec:spherical_harmonics}
The first modification we will discuss is a non-spherical perturbation induced by potential fluctuations that we express in terms of spherical harmonics. We consider this as a phenomenological modification that is however well-motivated by the observation that haloes in cosmological simulations do not form from truly spherical peaks. There is however another strong motivation to include such a perturbation: It is well known that spherically symmetric initial conditions, when evolved with a three-dimensional $N$-body code, undergo a strong radial orbit instability \citep[ROI; e.g.,][]{2006ApJ...653...43M, 2011MNRAS.414.3044V}. The ROI is always triggered numerically and often leads to a system whose major axis is aligned with a space-diagonal in tree-codes based on an oct-tree. While the instability is a real one, and an unstable system will of course undergo this instability, the fact that it is triggered by discreteness leads to a lack of numerical convergence. Convergence can be achieved however when the symmetry of the system is weakly broken already in the initial conditions. We discuss the effect of the ROI and resolved symmetry breaking in more detail in Appendix~\ref{subsec:spherical} and \ref{subsec:non-spherical}.

Motivated by these observations, we therefore consider a perturbation of the gravitational potential of the symmetric system, $\Phi_0(r)$, of the form
\begin{equation}
\Phi(r, \theta, \phi) = \sum_l \sum_m \frac{y_{lm}}{2l+1} \biggl (\frac{r}{r_{\rm p}} \biggr )^l \Phi_0(r) Y_{lm}(\theta, \phi), \label{eq:non_spherical_perturbation}
\end{equation}
where $Y_{lm}(\theta, \phi)$ are the spherical harmonics of degree $l$ and order $m$ ($|m| \leq l$).
The control parameters $y_{lm}$ and $r_{\rm p}$ determine the amplitude and radial scale of the perturbations.
Equation (\ref{eq:non_spherical_perturbation}) is of course motivated by a multipole expansion of the gravitational potential \citep[e.g.][]{2008gady.book.....B} around the proto-halo peak in the limit of $r \rightarrow 0$, i.e.,$\Phi(r, \theta, \phi) \propto r^l Y_{lm}(\theta, \phi)$ \citep[but see also][]{1992ApJ...386..375H}. 
In the simulations that we discuss in this paper, we set $r_{\rm p}=r_{\rm edge}/2$ where $r_{\rm edge}$ is the radius of the sphere of particles extracted from the perturbed particle lattice
\footnote{
The side length of the regular particle lattice before perturbing by the Zel'dovich approximation corresponds to $2 r_{\rm edge}$. 
}
, and $y_{00}=1$ (for the spherical term), $y_{22}=1/4$ and $y_{lm}=0$ for spherical harmonics with other $l$ or $m$. The axial ratio of the overdense part in the proto-halo patches is $\approx 1:0.93:0.87$ while the whole structure is almost spherical, i.e. the axial ratio of the whole patches is $\approx 1:1:1$. We note that interestingly our results do not strongly depend on the particular choice of $y_{lm}$, which we discuss in Appendix~\ref{subsec:yamp_shape}. 

\subsubsection{Modelling the graininess of progenitors and substructure}
\label{sec:gauss_noise}
An interesting question is whether the density profiles of DM haloes formed through gravitational collapse are robust to the dynamical impact of small-scale perturbations and minor mergers. Recently, \cite{2016MNRAS.461.3385O} and \cite{2016arXiv160403131A} demonstrated that the central density slope of the smallest microhaloes described as $\rho \propto r^{-1.5}$ gets shallower as the haloes grow through mergers \citep[see also][]{2014ApJ...788...27I}. They also showed that the central slope approaches the {\it universal} NFW-form, $\rho \propto r^{-1}$, and the central density structure becomes more resilient to the dynamical impact of subsequent mergers. \cite{2012MNRAS.427L..30P} investigated the impacts of small scale perturbations given in the form of sine waves and found that the perturbations can make the central density cusps shallower \citep[see also][]{1999ApJ...517...64H}. In the idealised $N$-body simulations of dissipationless collapse with the Gaussian random density field by \cite{2015ApJ...805L..16N}, the S{\'e}rsic profile naturally arises and the S{\'e}rsic index depends on the slope of the power spectrum of the Gaussian random field. 

\begin{figure}
  \centering 
   \includegraphics[width=85mm]{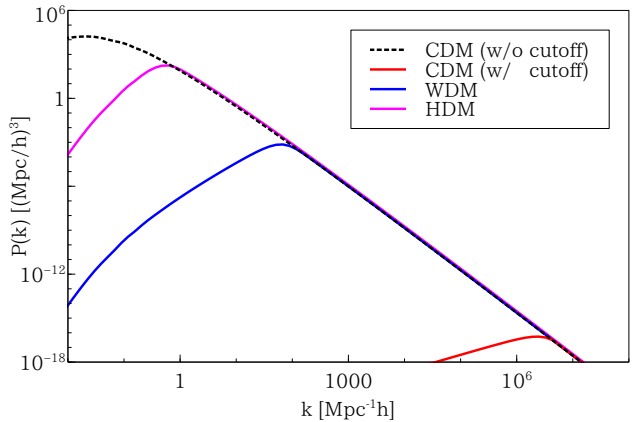}
   \caption{
     Power spectrum for the Gaussian noise, defined as $P_{\rm Gauss}(k) = P_{\rm w/o \ cut-off}(k) - P_{\rm w/ cut-off}(k)$, where $P_{\rm w/o \ cut-off}(k)$ and $P_{\rm w/ cut-off}(k)$ are the matter power spectra in the model of CDM without the cut-off (shown as black dotted line) and models with the cut-off. Red, blue and magenta lines represent the power spectra for the Gaussian noise in the models of CDM with the cut-off, of WDM and of HDM, respectively. 
       \label{fig:pk}
}
\end{figure}

Their results may explain why the NFW model is the universal form of the DM density profile in cosmological simulations in which DM haloes are perturbed by mergers. 
These observations motivate us to include a second additional ingredient: perturbations of the initial gravitational potential due to a Gaussian random field of density fluctuations. As we are not interested in major mergers here, we will consider random perturbations only on scales smaller than the scale of the free-streaming cut-off of each DM model (Table~\ref{tab:zi_zf}). We note that this aims to model, in a phenomenological way, the essential difference between cosmological simulations with a resolved cut-off and those with power all the way to the resolution limit of the simulation.
We therefore add a Gaussian random density perturbation field to the initial density field described by equation (\ref{eq:ini_tot}).
The power spectrum for the Gaussian density field, $P_{\rm Gauss}(k)$, is given as
\begin{equation}
P_{\rm Gauss}(k) = g_{\rm amp}[P_{\rm w/o \ cut-off}(k) - P_{\rm w/ cut-off}(k)], 
\end{equation}
where $k$ and $g_{\rm amp}$ represent the wavenumber and a parameter to control the amplitude of the Gaussian density perturbation. 
$P_{\rm w/o \ cut-off}(k)$ and $P_{\rm w/ cut-off}(k)$ are the matter power spectra in the model of CDM without the cut-off (shown as the black dotted line in Figure \ref{fig:pk}) and models with the cut-off. The coloured lines in Figure \ref{fig:pk} represent $P_{\rm Gauss}(k)$ in each model.

\subsection{Simulation code and numerical parameters}
\label{subsec:code_param}
Our $N$-body simulations are carried out with a tree code \citep{1986Natur.324..446B} to compute inter-particle forces and a Leapfrog scheme for time integration. The code we employ is designed to run on graphic processing unit (GPU) clusters. Following \cite{2011arXiv1112.4539N}, an oct-tree is built from particles on CPU cores and communicated with other processes, while the GPU cards subsequently evaluate the $N$-body interactions by traversing the tree \citep{2013JPhCS.454a2014O}. The softening length, $b$, is set to be $b=r_{\rm edge}/1280$, which corresponds to $\sim 10^{-3}r_{\rm vir}$ at the final redshift of the simulations.
\footnote{The softening length is kept constant in physical units throughout the simulations although cosmological simulations often keep it constant in comoving units.} 
The opening angle of the tree algorithm is set to $\theta = 0.3$ in all runs. We note that this value of $\theta$ is smaller, i.e. higher force accuracy, than the value typically adopted in cosmological simulations, $\theta=0.6-0.7$ \citep[e.g.][]{2003MNRAS.338...14P}, but is required to achieve sufficient force accuracy in the simulations (for details see Appendix \ref{subsec:param_choice}). 
It might be possible to achieve similar accuracy with higher $\theta$ but with a higher order multipole expansion of the tree nodes (which are accurate up to dipole order in our implementation).
We also note that we adopted a traditional tree algorithm in our simulations, while the tree particle mesh scheme is commonly used in recent cosmological simulations and the errors in simulations could be different even if the same $\theta$ and multipole expansion of the same order are adopted. 
We employ 8,680,336 particles in each run and $\sim$ 1 million particles are contained in the virial radius of the DM halo at the end of the simulations. The simulation results appear numerically converged at the scales we consider (cf. Appendix~\ref{subsec:non-spherical}). 

We use the definition of virial overdensity of \cite{1998ApJ...495...80B} throughout this paper when we refer to virial radius and mass.
The potential minimum in the patch is defined as the halo centre and the virial mass and radius are computed with the spherical-overdensity method \citep[e.g.][]{1974ApJ...187..425P}.

\section{Results}
\label{sec:results}
In this section, we present the results of our numerical simulations. We will try to discern the relative contributions of (1) the presence of an initial core in the proto-halo profile, (2) the slope of the proto-halo profile, and (3) the role of small-scale noise, on the final profile of the collapsed DM halo. We have relegated additional material demonstrating numerical convergence of our results as well as the impact of certain parameter choices to the appendix.

\subsection{Visual impression: role of core and noise}
\begin{figure*}
  \centering 
   \includegraphics[width=155mm]{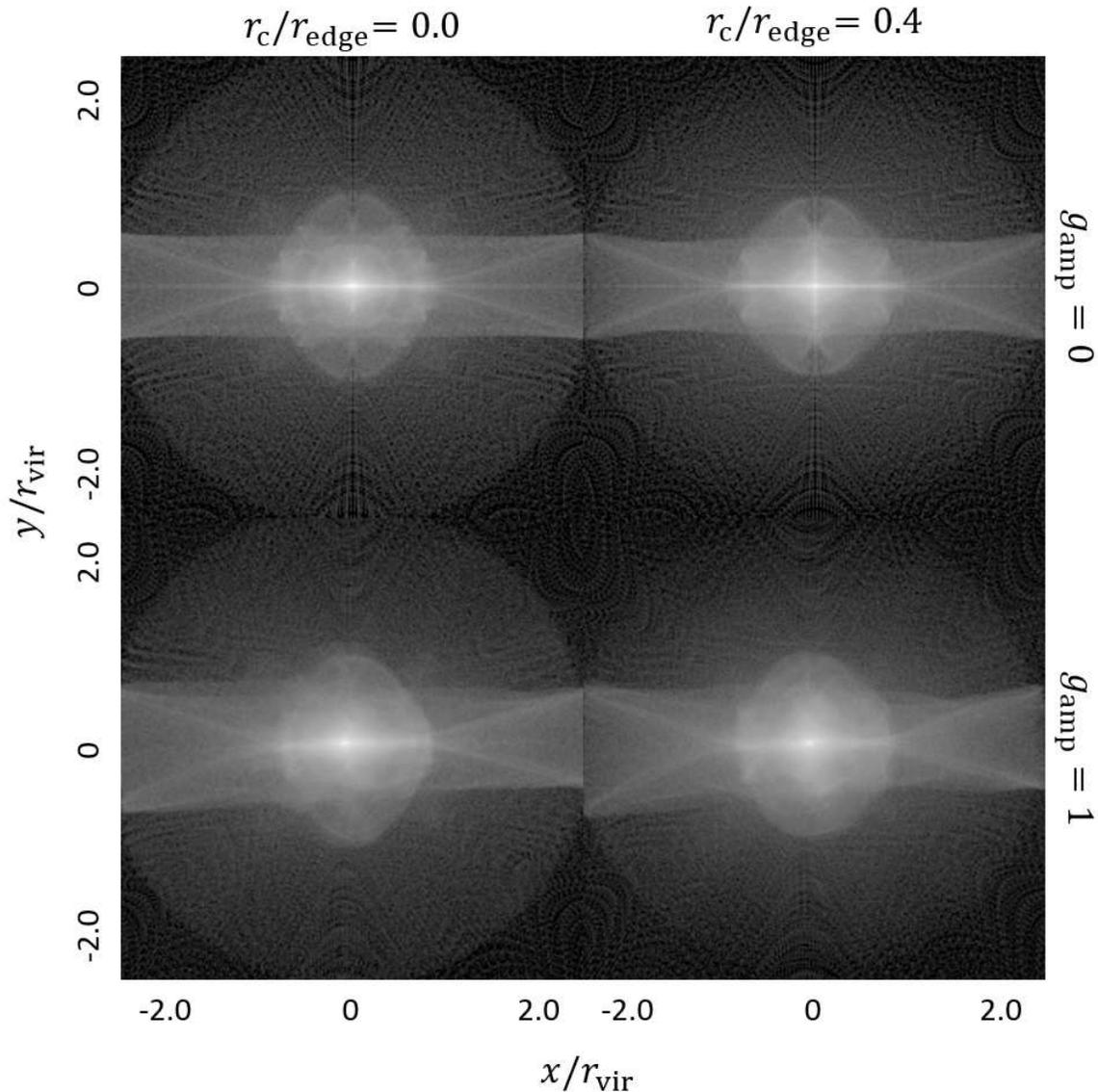}
   \caption{
     Column density map of microhaloes ($\epsilon=0.05$) at $z=31$. 
     Left and right columns demonstrate the results of simulations of $r_{\rm c}/r_{\rm edge}=0.0$ and 0.4. 
     Panels in the upper and lower rows illustrate the images in simulations of $g_{\rm amp}=0$ and 1, respectively. 
     Symmetric structures have been formed because of the non-spherical perturbation. 
     The global structure in simulations of $g_{\rm amp}=0$ is symmetric and almost independent of $r_{\rm c}/r_{\rm edge}$ though smaller caustics are seen in the run of $r_{\rm c}/r_{\rm edge}=0.0$. 
     The caustics in the lower panels are less sharp than those in the upper panels because of the Gaussian noise.
     The noise distorts the global structure as well. 
     \label{fig:image_rc_gamp}
}
\end{figure*}

Before we discuss the results of our collapse simulations in more detail, in Figure \ref{fig:image_rc_gamp}, we give a visual impression of the density structure of a DM halo originating from an initially weakly aspherically perturbed proto-halo profile. The two panels in the left column show haloes resulting from proto-haloes without an initial core (i.e. with power-law asymptotics as $r\to 0$), while those in the right column are the results of initially cored proto-halo profiles. The top row panels, on the one hand, show the haloes forming from an initial perturbation with no small-scale Gaussian noise, while the bottom row shows the same haloes when the initial proto-halo density profiles were perturbed with small-scale Gaussian noise.
The slope $\epsilon$ of the proto-halo profile was set here to that corresponding to microhaloes ($\epsilon=0.05$).
One clearly sees that sharp, highly symmetric caustics have formed during gravitational collapse in the runs without the Gaussian noise (upper panels). 
We observe that the global structure does not change significantly even if the size of the initial core is varied (left vs. right panels), while sharper caustics are formed in the centre of the halo in the run without an initial core (i.e. $r_{\rm c}/r_{\rm edge}=0.0$). In all cases, the Gaussian noise makes the caustics less sharp and weakly distorts the global halo shape (lower panels). We also emphasise once more that without the weak global aspherical perturbation, we observe a strong radial orbit instability of numerical origin that prevents a clean convergence (see Appendix~\ref{subsec:spherical}).

\subsection{Impact of the proto-halo core}
\label{subsec:ini_core}

\begin{figure}
  \centering 
   \includegraphics[width=85mm]{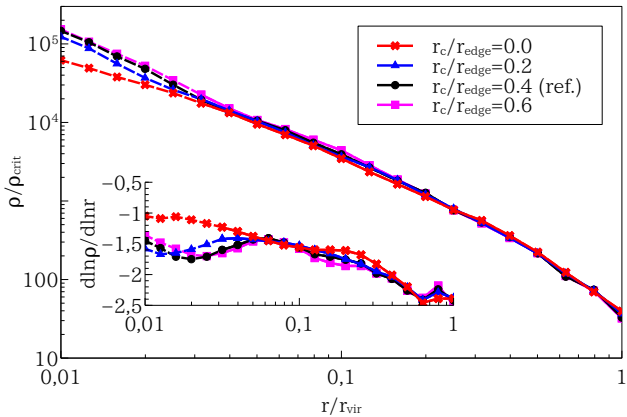}
   \caption{
     Radial density profile of microhaloes ($\epsilon=0.05$) at $z=31.0$ in the runs varying the size of the initial density cores, $r_{\rm c}$. 
     The inset panel shows the profile of the logarithmic density slope.
     The slope at a radial bin is estimated from the power-law fitting for the density profile with six adjacent radial bins. 
     The condition, $\tau_{\rm rel}(r) \geq \tau_{\rm sim}$, is (not) satisfied in the radial range of solid (dashed) lines. 
     Red, blue, black and magenta lines show the results of the runs with $r_{\rm c}/r_{\rm edge}=0.0$, 0.2, 0.4 and 0.6, where $r_{\rm edge}$ is the truncation radius of the initial particle distribution. 
     Steeper density cusps are formed and higher central densities are obtained in the runs with the larger $r_{\rm c}$. 
     This trend almost saturates when $r_{\rm c}/r_{\rm edge} \geq 0.4$. 
     Hereafter we adopt $r_{\rm c}/r_{\rm edge} = 0.4$ in the runs to study the halo formation near the free-streaming scale. 
     \label{fig:rc_dens}
}
\end{figure}

As a first step in our analysis, we wish to investigate the role of the shape of the proto-halo peak on the final halo density profile. In our model, as described above, the peak shape is essentially modelled by two parameters, the slope $\epsilon$ as well as the presence or absence of an initial core due to a cut-off in the spectrum. 
In the first set of runs, we fix the parameter to control the initial density slope to be $\epsilon=0.05$ (corresponding to the scale of microhaloes), and vary the core size in the initial proto-halo profile. 
The run with $r_{\rm c}/r_{\rm edge} = 0.0$ is a simple model for the gravitational collapse in cosmological simulations without the cut-off in the matter power spectrum where the resulting DM haloes are routinely found to have an NFW-like central cusp, i.e. $\alpha \sim 1.0$. 
Other runs with $r_{\rm c}/r_{\rm edge} > 0.0$ model the gravitational collapse in cosmological simulations with the cut-off.

Figure \ref{fig:rc_dens} presents the spherically averaged density profile, $\rho(r)$, and the inset shows the profile of the logarithmic density slope, $d\ln{\rho} / d\ln{r}$, respectively. We indicate the radius up to which two-body relaxation might have possibly affected our results by the transition from solid to dashed lines: 
the radial bins at which the condition $\tau_{\rm rel}(r) \geq \tau_{\rm sim}$ is (not) satisfied are represented by solid (dashed) lines. Here $\tau_{\rm sim}$ is the simulation run-time to the output, and $\tau_{\rm rel}(r)$ is the time-scale of two-body relaxation, defined as \citep[cf.][]{2003MNRAS.338...14P}
\begin{equation}
\tau_{\rm rel}(r) = \frac{5}{3} \frac{N(r)}{8 \ln{N(r)}} \sqrt{\frac{r^3}{GM(r)}}, 
\end{equation}
where $N(r)$ and $M(r)$ are the number of particles and mass enclosed within the radius $r$ respectively.  
The density slope in the runs of $r_{\rm c}/r_{\rm edge} = 0.4$ (black) and 0.6 (magenta) approaches $d\ln{\rho}/d\ln{r} \approx -1.5$ at the centre of the collapsed systems. 
This is consistent with the results of cosmological $N$-body simulations for microhaloes \citep[e.g.,][]{2014ApJ...788...27I, 2016arXiv160403131A}. 
As expected, the NFW-like central cusp, $d\ln{\rho}/d\ln{r} \approx -1.0$, is obtained in the run with $r_{\rm c}/r_{\rm edge} = 0.0$ (red).

The central densities of the collapsed systems which had the larger $r_{\rm c}$ in the initial state appear expectedly higher than the others. {\em Clearly, the one run starting from the pure power-law profile is qualitatively different from all other runs that started with an initial core.}
On the other hand, the density structure in the outskirts is virtually the same in all cases. 
The slope in the outskirts ($r/r_{\rm vir} > 0.06$) is almost the same in all runs as well and the central cusp in systems with larger $r_{\rm c}$ is steeper, but the increase of the central density and steepening of the slope almost saturate for $r_{\rm c}/r_{\rm edge} \geq 0.4$. 
Due to this behaviour, we employ $r_{\rm c}/r_{\rm edge}=0.4$ as the reference model for the cases with the cut-off in what follows, and restrict our analysis to only the comparison between runs with and without an initial proto-halo core.

The wiggles in the centre of DM haloes seen in the slope profile might not usually be resolved in cosmological simulations. As mentioned in Section \ref{subsec:code_param}, the opening angle of the tree algorithm adopted in our simulations, $\theta=0.3$, is smaller than the values typically adopted in cosmological simulations, $\theta=0.6-0.7$, and hence our simulations would have higher force accuracy. The opening angle of $\theta=0.3$ is carefully chosen to achieve numerical convergence and the wiggle disappears when the typical value, $\theta=0.6$, is adopted (see Appendix \ref{subsec:non-spherical} and \ref{subsec:param_choice} for details).

It is also worth noting that the haloes are well relaxed at $z_{\rm f}$. We computed the virial ratio, $2K/|U|$, of DM haloes in the two runs of $r_{\rm c}/r_{\rm edge}=0.0$ and 0.4, where $K$ and $U$ are the total kinetic and potential energy of the haloes, respectively. Their virial ratio stays $\sim 1$ after $z=40$ in the runs. The time interval between $z=40$ and $z_{\rm f}=31$ corresponds to $\sim 3.8$ times the free-fall time measured at the virial radius at $z=z_{\rm f}$. The stability of the density profile is also shown in the lower panels of Figures \ref{fig:bino_gaussian_rc0.40} and \ref{fig:bino_gaussian_rc0.00} (see dashed and thick solid lines). 

\begin{figure}
 \centering 
  \includegraphics[width=85mm]{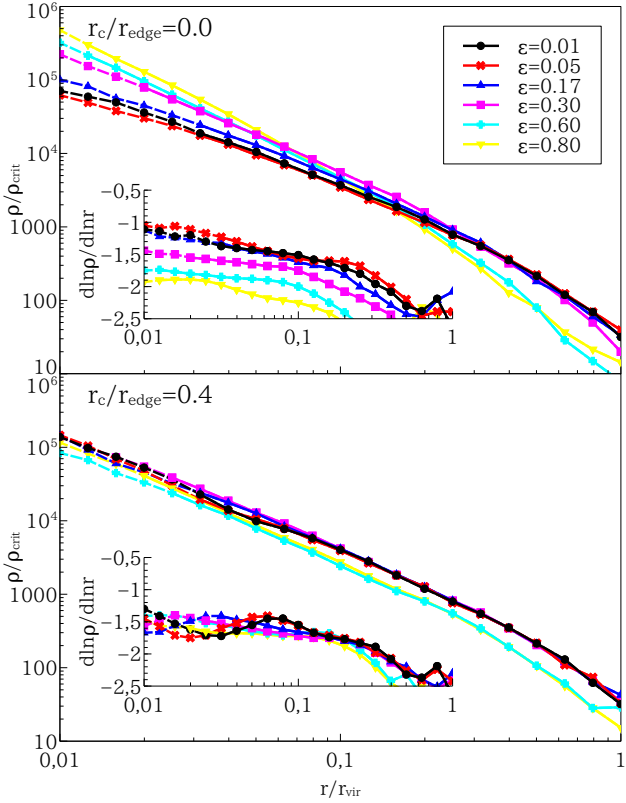}
   \caption{
     Radial density profile of haloes in runs varying the parameter to control the initial slope, $\epsilon$. 
     The inset panel shows the profile of the logarithmic density slope. 
     Snapshots at the redshift to finish the simulations, $z_{\rm f}$, are shown (see Table \ref{tab:zi_zf}). 
     Upper and lower panels present the results from the runs of $r_{\rm c}/r_{\rm edge}=0.0$, the models without the cut-off in the initial power spectrum, and of 0.4, the models with the cut-off. 
     In each panel, black, red, blue, magenta, cyan and yellow lines represent the results of runs of $\epsilon=0.01, 0.05, 0.17, 0.30, 0.60$ and 0.80, respectively. 
     The condition, $\tau_{\rm rel}(r) \geq \tau_{\rm sim}$, is (not) satisfied in the radial range of solid (dashed) lines. 
     The results of the runs of $r_{\rm c}/r_{\rm edge}=0.0$ well match the analytical prediction by L06. 
     In the runs of $r_{\rm c}/r_{\rm edge}=0.4$, the central density slope is almost independent of $\epsilon$ and is $\sim -1.5$. 
     \label{fig:eps_dep}
}
\end{figure}

In the next step, we study how the final density structure depends on the density slope of the proto-halo profile. 
Figure~\ref{fig:eps_dep} shows the radial density profile of haloes in runs in which we vary the parameter $\epsilon$: black, red, blue, magenta, cyan and yellow lines show the results of runs with $\epsilon=0.01, 0.05, 0.17, 0.30, 0.60$ and 0.80, respectively. 
The resulting central density slope in the runs without initial core (i.e. $r_{\rm c}/r_{\rm edge}=0.0$; upper panel) is $d\ln{\rho}/d\ln{r} \approx -1.0$ for $\epsilon \la 0.2$, but shows a monotonous increase with $\epsilon$ if $\epsilon \ga 0.2$. These simulation results match well the theoretical predictions by L06. The final halo profile is radically different however for the runs with an initial density core. Despite $\epsilon$ varying by almost two orders of magnitude, the central density slope in the runs with initial cores ($r_{\rm c}/r_{\rm edge}=0.4$; bottom panel) appears insensitive to $\epsilon$ and is $d\ln{\rho}/d\ln{r} \sim -1.5$, universally. 
We have combined these results (along with others that we present below) into {\em an overview of the relation between initial and final slope, with and without initial proto-halo core, in Figure \ref{fig:lu_diagram}}, which can be thought of as a visual summary of our findings.

\subsection{Impact of small-scale Gaussian noise}
\label{subsec:noise}

In cosmological $N$-body simulations with the cut-off in the initial matter power spectrum, the central density cusps of $\alpha > 1$ are seen in microhaloes ($\epsilon \approx 0.05$, \citealp[][]{2010ApJ...723L.195I, 2013JCAP...04..009A, 2014ApJ...788...27I, 2016arXiv160403131A}). Similar simulations for WDM ($\epsilon \approx 0.17$, e.g. \citealp{2001ApJ...559..516A, 2007ApJ...665....1B, 2014MNRAS.439..300L}, but see also \citealp{2015MNRAS.450.2172P}) or HDM haloes ($\epsilon \approx 0.60$, e.g. \citealp{2009MNRAS.396..709W}) found that the density structure of the haloes is well described by the NFW model ($\alpha=1$) although they are expected to be formed through monilithic collapse just like microhaloes. 
On the other hand, density cusps of $\alpha \sim 1.5$ are readily formed in all our simulations when an initial core is present ($r_{\rm c}/r_{\rm edge}=0.4$) for a wide range of power-law profiles above the cut-off scale, $0.01 \leq \epsilon \leq 0.8$. 

What might cause this apparent difference? In the monolithic collapse simulations that we have discussed above, the haloes form through the smooth accretion of DM. In cosmological simulations, however, the dynamical impact of interactions with nearby systems plays a key role to determine the density structure of DM haloes. 
Even if the smallest DM haloes have central cusps of $\alpha = 1.5$, the density cusp becomes increasingly shallower as the haloes undergo subsequent mergers \citep[cf.][for the effect of mergers on the evolution of microhalo profiles]{2016MNRAS.461.3385O, 2016arXiv160403131A}. 
These simulations also showed that after reaching $\alpha \sim 1$, i.e. the NFW profile, the central cusp becomes more resilient against mergers, and thus the resilience may explain why the NFW model is the {\it universal} DM density profile.

It is thus entirely possible that previous studies based on cosmological simulations of WDM or HDM cosmology may have focused on the density structure of DM haloes whose central cusp has already been transformed into that of the NFW model through mergers. 
For example, \cite{2001ApJ...559..516A}, \cite{2007ApJ...665....1B} and \cite{2014MNRAS.439..300L} analysed Milky Way sized or larger WDM haloes, $\ga$1000 times more massive than the mass scale of the smallest WDM haloes. And although \cite{2001ApJ...559..516A} also showed the density profile of smaller WDM haloes with mass of $\sim 10^9 M_{\rm \odot}$, the spatial resolution was not enough to resolve down to $\sim1 \%$ of the virial radius of the haloes.

While there is thus strong evidence for an evolution of the density profiles as haloes grow in mass by mergers, one might wonder to what degree the presence of any small scale graininess might drive such a process. Cosmological simulations with a cutoff-spectrum should not produce haloes below the mass-scale associated with that scale. The presence of spurious haloes due to artificial fragmentation in such simulations and the associated difficulties to make accurate predictions of the low-mass halo mass function has been discussed at length in the literature \citep[e.g.,][and references therein]{2013MNRAS.434.3337A, 2014MNRAS.439..300L} and briefly summarised in the introduction to this article. It is thus not unreasonable to suspect that when spurious haloes merge with physical ones, the density structure of physical DM haloes may be altered. 
In addition, they can disturb the formation of physical ones by capturing mass which should be contained in a physical one. 

The sharpness of the cut-off in the power spectrum may also play a role to cause the difference between the density structures of micro- and WDM or HDM haloes. The cut-off imposed in the scale of microhaloes \citep{2004MNRAS.353L..23G} is sharper than those in the scales of the smallest WDM or HDM haloes \citep{2001ApJ...556...93B}. Because of the sharp cut-off, microhaloes may form and stabilize before other smaller systems, including spurious haloes, affect them.

\begin{figure}
  \centering 
   \includegraphics[width=85mm]{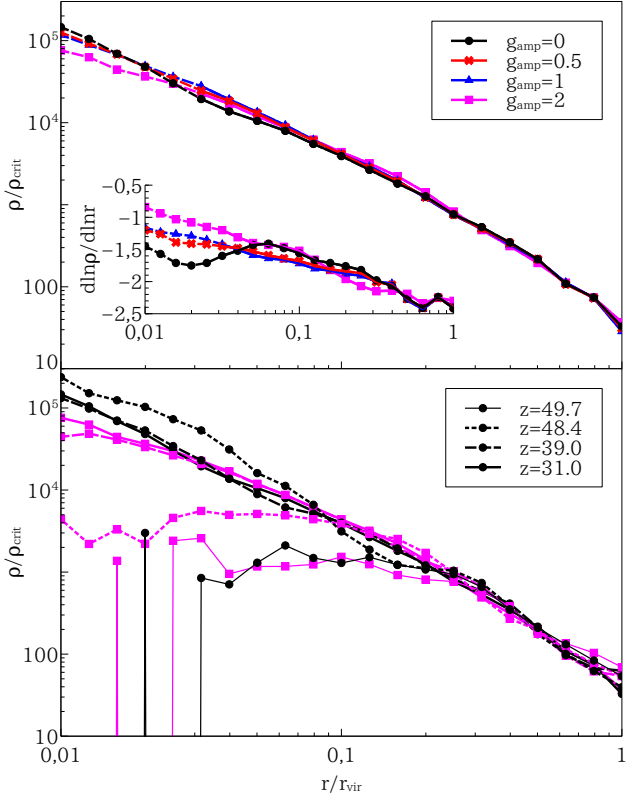}
   \caption{
     Impacts of the Gaussian noise onto the density structure of CDM haloes in the scale of microhaloes.
     The initial overdense patch is characterized as $r_{\rm c}/r_{\rm edge}=0.4$ and $\epsilon=0.05$. 
     {\it Upper panel:} Radial density structure of haloes obtained from runs varying the amplitude of the Gaussian noise, $g_{\rm amp}$.
     Black, red, blue and magenta lines represent the results of runs with $g_{\rm amp}=0, 0.5, 1$ and 2, respectively. 
     In runs with higher $g_{\rm amp}$, shallower cusps are formed. 
     {\it Lower panel:} Evolution of the density structure. 
     Thin solid, dotted, dashed and thick solid lines are the snapshots at $z=49.7$, 48.4, 39.0 and 31.0, respectively. 
     The formation of the central cusp is suppressed by the noise in the run of $g_{\rm amp}=2$ (magenta), and the halo eventually has the shallower cusp than that in the run without the noise ($g_{\rm amp}=0$; black). 
     \label{fig:bino_gaussian_rc0.40}
}
\end{figure}

In this subsection, we study the impact of a grainy density field on the formation of our idealised haloes. To this end, we present additional simulations in which we perturb the initial particle lattice with Gaussian random density perturbations. Since we do not want to induce large scale perturbations to the proto-halo profile, we restrict the perturbation spectrum to scales smaller than the cut-off scale. 
In the upper panel of Figure \ref{fig:bino_gaussian_rc0.40}, we show the density profile of haloes in runs which model the halo formation in cosmological simulations with the cut-off ($r_{\rm c}/r_{\rm edge}=0.4$), varying the amplitude of the Gaussian noise, $g_{\rm amp}$.
In the runs with higher $g_{\rm amp}$, the central density decreases and the density slope at the centre becomes shallower compared to the runs with lower $g_{\rm amp}$. 
The lower panel of Figure~\ref{fig:bino_gaussian_rc0.40} clearly shows that the formation of the central cusp is disturbed by the noise in the run with $g_{\rm amp}=2$ (magenta), and the halo eventually has a significantly shallower cusp than that in the run without the noise ($g_{\rm amp}=0$; black). 
We note that the formation of the central cusp is disturbed by the noise (see lines of the same line style), but the noise does not work to flatten the cusp once it has been formed. 
Supposing that the impact of spurious haloes are mimicked by the Gaussian noise, it is thus entirely plausible that one of the reasons to find the NFW halo profiles in cosmological simulations of WDM and HDM cosmologies may be driven by small-scale noise and spurious haloes if the cut-off scale is not very well resolved.

\begin{figure}
  \centering 
   \includegraphics[width=85mm]{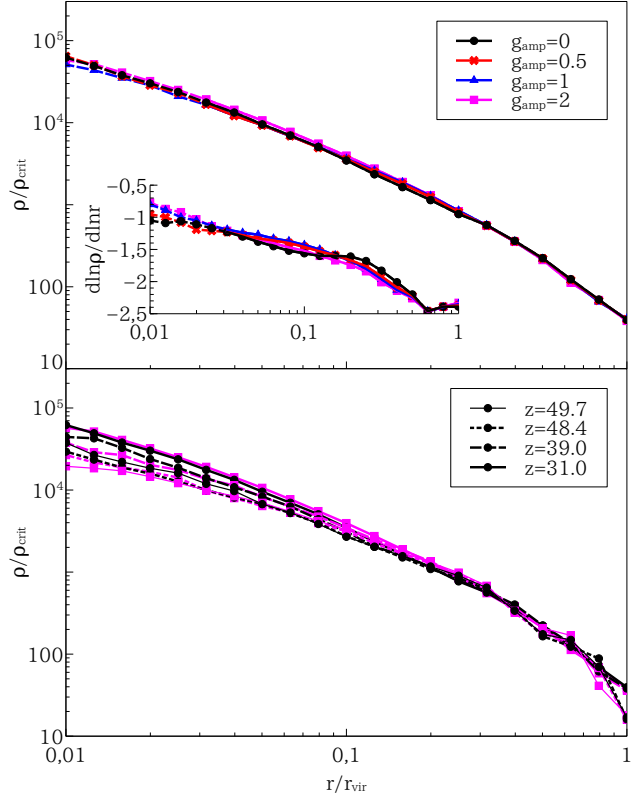}
   \caption{
     Impacts of the Gaussian noise onto the density structure of CDM haloes in the scale of microhaloes.
     The initial overdense patch is characterized as $r_{\rm c}/r_{\rm edge}=0.0$ and $\epsilon=0.05$. 
     {\it Upper panel:} Radial density structure of haloes obtained from runs varying the amplitude of the Gaussian noise, $g_{\rm amp}$.
     Black, red, blue and magenta lines represent the results of runs with $g_{\rm amp}=0, 0.5, 1$ and 2, respectively. 
     In runs with higher $g_{\rm amp}$, shallower cusps are formed, but the impact in the slope is smaller than that shown in Figure \ref{fig:bino_gaussian_rc0.40}. 
     {\it Lower panel:} Evolution of the density structure. 
     Thin solid, dotted, dashed and thick solid lines are the snapshots at $z=49.7$, 48.4, 39.0 and 31.0, respectively.
     The formation of the central cusp is suppressed as with the run in Figure \ref{fig:bino_gaussian_rc0.40}, but the impacts is weaker. 
     \label{fig:bino_gaussian_rc0.00}
}
\end{figure}

Figure \ref{fig:bino_gaussian_rc0.00} shows the resultant density structure of DM haloes obtained in runs which model the halo formation in cosmological simulations without the cut-off ($r_{\rm c}/r_{\rm edge}=0.0$).
The formation of the central cusp is disturbed and the formed central cusp in the runs of the higher $g_{\rm amp}$ is shallower than that in the runs of the lower $g_{\rm amp}$, as presented in Figure \ref{fig:bino_gaussian_rc0.40}, but the impacts is much less significant compared to the runs with the cut-off because in this case the ratio of the amplitude of the density perturbation of the noise to that of the smooth density peak would be smaller than that in the case with the cut-off. 
Low-mass haloes formed in the runs would correspond to subhaloes in cosmological simulations without the cut-off and the simulation results indicate that the NFW profile appears robust against the dynamical impacts caused by substructures and other small-scale noise.
Hence once the central cusps of $\alpha=1$ is built up, the density structure would be retained for a long time and seen as the universal form in cosmological simulations. 

One thing, which our simplified simulations do not take into account, is the tidal torque of filaments and larger cosmic structures which might deform overdense patches \citep[e.g.][see also BBKS]{1970Afz.....6..581D,1984ApJ...286...38W, 2002MNRAS.332..325P} in more complicated ways than modelled by our non-spherical perturbation. 
Because of that, DM haloes have a small net angular momentum which might also play a role in altering their density structure \citep[e.g.][]{2010PhRvD..82j4044Z}. On the other hand, \cite{2016ApJ...822...89L} presented that in most of the cases they studied, the NFW profile arises even if the amount of angular momentum of the proto-halo patches is significantly varied. We note that our global perturbation does not induce a net angular momentum, although individual particles of course have non-zero angular momentum. The role of non-zero global angular momentum is an aspect which deserves further investigation, but is beyond the scope of our present study.

\subsection{Overview of all simulations}
\begin{figure*}
  \centering 
  \includegraphics[width=120mm]{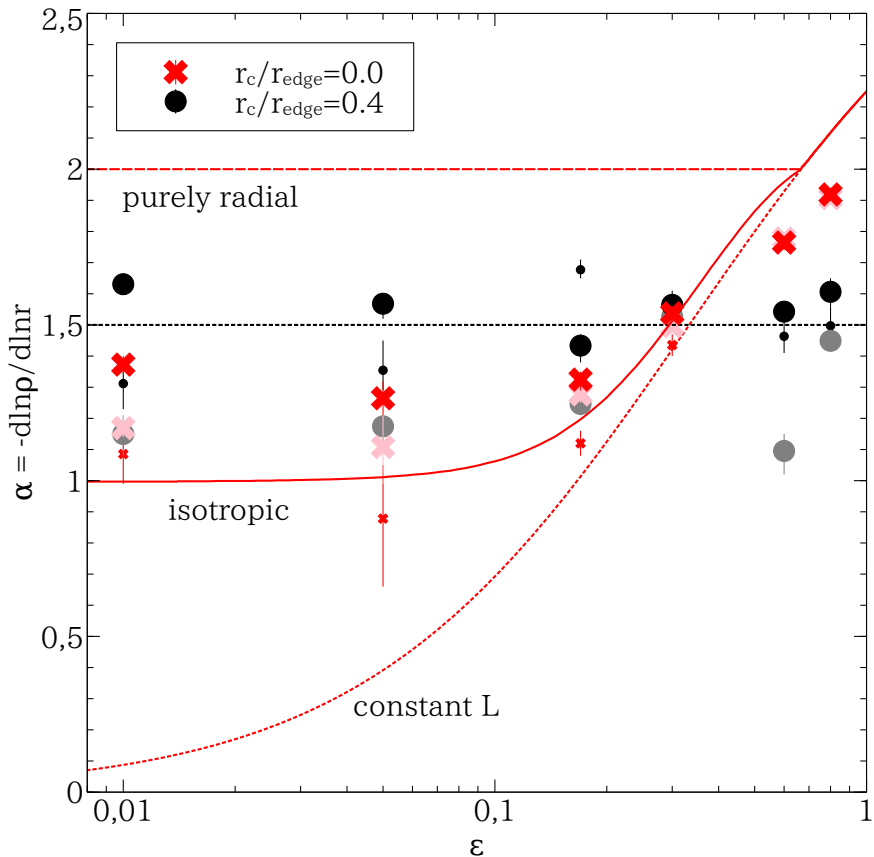}
   \caption{
     The logarithmic density slope, $\alpha = -d\ln{\rho}/dr$, as a function of the parameter to control the slope of the proto-halo patch, $\epsilon$.
     Red dashed, dotted and solid curves show the models with the purely radial orbits \citep{1984ApJ...281....1F}, with the fixed ${\cal L}$ for all mass shells where ${\cal L} \equiv L/L_{\rm c}$ is the ratio between the angular momentum, $L$, and that of the circular orbit, $L_{\rm c}$ \citep{2001MNRAS.325.1397N}, and with the isotropic velocity structure (L06), respectively. 
     The three curves overlap each other for $\epsilon \geq 2/3$. 
     Black dotted line corresponds to $\alpha = 1.5$. 
     Crosses and circles are the density slope measured in the simulations of $r_{\rm c}/r_{\rm edge}=0.0$ and 0.4.
     The averaged $\alpha$ values of the last 11 snapshots are shown and error bar demonstrates the 1st and 11th values.  
     Red and black symbols represent the results of simulations without the noise ($g_{\rm amp}=0$). 
     For the large symbols, the inner slope is measured at $0.1r_{-2}$, where $r_{-2}$ is the radius at which the logarithmic density slope is -2, except for runs with $r_{\rm c}/r_{\rm edge}=0.0$ and $\epsilon=0.60$ or 0.80 because the condition of two-body relaxation is not satisfied at $0.1r_{-2}$ (it is satisfied in the other runs). 
       Hence we measure the inner slope at the innermost radial bin to satisfy the condition (at $\sim 0.15$ and $0.3 r_{\rm -2}$ in the runs of $\epsilon=0.60$ and 0.80, respectively). 
       The small symbols represent the inner slope measured at $0.01r_{\rm vir}$. Although the condition of two-body relaxation is not satisfied at the radius, we show them for comparison. 
     Small red ones follow red solid line well and black ones show $\alpha \sim 1.5$ in the parameter range we study, $0.01 \leq \epsilon \leq 0.8$. 
     Pink and grey ones are obtained in the simulations of $g_{\rm amp}=2$ ($\alpha$ is measured at $0.1r_{-2}$). 
     Comparing grey symbols with black ones, the noise makes the central cusp shallower. 
     Cusps formed in the runs of $r_{\rm c}/r_{\rm edge}=0.0$ are more robust to the noise (see red and pink symbols). 
     \label{fig:lu_diagram}
}
\end{figure*}

In order to facilitate the comparison of our various simulation results, we next show the essence of our study in a single graph. Figure~\ref{fig:lu_diagram} summarises the central density slope, $\rho \propto r^{-\alpha}$, as a function of the parameter, $\epsilon$, describing the proto-halo power-law profile (on scales larger than the cut-off, if one is present). The red and black dotted lines represent the predictions by analytical studies and $\alpha=1.5$, respectively. Red crosses show the logarithmic density slope measured in the simulations without initial core, i.e. $r_{\rm c}/r_{\rm edge}=0.0$, while black circles show the corresponding results for the runs with an initial core, i.e. $r_{\rm c}/r_{\rm edge}=0.4$.
For the large symbols, the inner slope is measured at $0.1r_{-2}$, where $r_{-2}$ is the radius at which the logarithmic density slope is -2, or at the innermost radial bin to satisfy the condition of two-body relaxation. We also show the slope measured at $0.01r_{\rm vir}$ with the small symbols for comparison although the condition of two-body relation is not satisfied at the radius. Small red ones follow red solid line well and black ones show $\alpha \sim 1.5$ in the parameter range we study, $0.01 \leq \epsilon \leq 0.8$.

The pink crosses and grey circles show the results {\em including} Gaussian small-scale noise ($g_{\rm amp}=2$). 
We note that both the red and pink crosses roughly follow the relation of L06, given by the red solid line, over the entire range of $0.01 \leq \epsilon \leq 0.8$ that we studied. \cite{2011MNRAS.414.3044V} studied a similar model over the smaller parameter range $0.4 \leq \epsilon \leq 0.8$, where a strong relation between $\alpha$ and $\epsilon$ exists, and obtained results consistent with ours. While the runs without a cut-off are thus remarkably insensitive to Gaussian noise (see red and pink crosses and also Figure \ref{fig:bino_gaussian_rc0.00}), the situation for the models with a cut-off is dramatically different. The unperturbed initially cored models, shown by the black circles, follow an almost constant $\alpha \sim 1.5$ over the entire range of $0.01 \leq \epsilon \leq 0.8$. Once perturbations are added, the profiles all become shallower (see black and grey circles, and also Figure~\ref{fig:bino_gaussian_rc0.40}) and for $\epsilon\lesssim  0.3$ consistent with the relation of L06 and therefore all the other profiles.

\section{Phenomenological Model for the $\alpha=1.5$ structure}
\label{sec:phenom_model}

As demonstrated in the previous section, we robustly obtained steep central cusps of $\alpha \sim 1.5$ when the simulations start from a cored proto-halo profile and no small-scale noise is included. This is in good agreement with previous studies of microhaloes in cosmological simulations. The cuspy proto-halo profiles in contrast lead to shallower central cusps for small $\epsilon$ and the dependence on larger values of $\epsilon$ as predicted by previous analytical studies. The persistence of the $\alpha\sim 1.5$ cusps for cored proto-haloes requires however further understanding, so that we will give some phenomenological motivation in what follows.

Let us consider the radial motion of a mass shell of initial radius, $r_{\rm i}$. The specific energy of the mass shell, $e$, can be written as
\begin{eqnarray}
  e &=& \frac{H^2(z_{\rm i}) r_{\rm i}^2}{2} + \frac{L^2}{2r_{\rm i}^2} - \frac{G M_0(<r_{\rm i})}{r_{\rm i}} - \frac{G\delta M_{\rm i}}{r_{\rm i}} \nonumber \\
    &=& \frac{L^2}{2r_{\rm i}^2} - \frac{G\delta M_{\rm i}}{r_{\rm i}}, \label{eq:e_cons}
\end{eqnarray}
where $H(z_{\rm i})$ and $L$ are the Hubble constant at the initial redshift, $z_{\rm i}$, and the specific angular momentum of the shell.
The background mass enclosed within $r_{\rm i}$ is given as $M_0(<r_{\rm i})=(4\pi/3)\rho_{\rm crit}(z_{\rm i}) r_{\rm i}^3$ where $\rho_{\rm crit}(z_{\rm i})$ is the critical density of the universe at $z_{\rm i}$ and $\delta M_{\rm i}$ means the perturbation mass. 
The first and third terms in the first row of equation (\ref{eq:e_cons}) cancel out each other by definition in the Einstein-de~Sitter model which provides a good approxomation for the flat universe model at high redshifts. 
Assuming conservation of the specific energy and angular momentum then we expect the following relation to hold before shell-crossing: 
\begin{equation}
  \frac{L^2}{2r_{\rm i}^2} - \frac{G\delta M_{\rm i}}{r_{\rm i}} = \frac{v_{\rm r}^2}{2} + \frac{L^2}{2r^2} - \frac{G\delta M_{\rm i}}{r}, 
\end{equation}
where $v_{\rm r}$ is the radial velocity of the shell when it is at $r$. 
The conserved specific angular momentum of the shell is given by 
\begin{equation}
  L^2 = 2 {\cal L}^2_{\rm i} G \delta M_{\rm i} r_{\rm i} = 2 {\cal L}^2 G \delta M_{\rm i} r, \label{eq:l_cons}
\end{equation}
where ${\cal L}_{\rm i}$ and ${\cal L}$ are the ratios between the specific angular momentum and that of the circular orbit at $z_{\rm i}$ and when the shell is at $r$.
In the cases we study in this paper, ${\cal L}_{\rm i}$ would be much smaller than 1 because the initial particle orbits are almost purely radial. 
Then, the radial velocity is given as
\begin{equation}
  v_{\rm r} = \pm \sqrt{2 G \delta M_{\rm i} \biggl [ \frac{1-{\cal L}^2}{r} - \frac{1-{\cal L}^2_{\rm i}}{r_{\rm i}} \biggr ]}. 
\end{equation}
Though equation (\ref{eq:l_cons}) implies that ${\cal L}^2 \propto r^{-1}$, it does not exceed 1 by definition ($0 \leq {\cal L} \leq 1$) and the radial dependence of $v_{\rm r}$ would be
\begin{equation}
  v_{\rm r} \propto r^{-0.5}. \label{eq:vr}
\end{equation}
The time of the shell to spend at $r$, $\delta t|_r$, is 
\begin{equation}
  dt|_r = \frac{\delta r}{v_{\rm r}} \biggl |_r \propto r^{1.5}  
\end{equation}
and is expected to be proportional to the mass profile, $M(r)$. 
Hence the radial dependence of the density structure is $\rho \propto r^{-1.5}$. 
We note that, e.g. \cite{1985ApJS...58...39B} derived the same radial dependence for the density structure of pressureless fluids collapsing onto a central object (see their section 2). In fact, $r^{-1.5}$ profiles have commonly been predicted in the classical proto-stellar collapse studies, where the density profile of an unstable isothermal sphere is predicted to have a slope of $-1.5$ due to free fall of the material interior to an outward travelling 'rarefaction wave' (see e.g. \citealp{1977ApJ...214..488S,1977ApJ...218..834H,1985MNRAS.214....1W}, but see also the Larson-Penston solution by \citealp{1969MNRAS.145..271L} and \citealp{1969MNRAS.144..425P}).

However, such free fall solutions might not be valid for the longer time evolution of collisionless systems after shell crossing. After the first shell crossing, the collapsed systems would approach to the dynamically relaxed state from the centre where the dynamical timescale is shorter than that measured at the outskirts and the system is expected to have the density structure of $\rho \propto r^{-1.5}$, at least for a while, because the infalling material is supplied from the outskirts. Hydrostatic equilibrium between gravity and velocity dispersion, $\sigma$, requires that
  \begin{equation}
    \frac{G M(r)}{r^2} = -\frac{1}{\rho} \frac{d (\rho \sigma^2)}{dr}. \label{eq:hydro_sta}
  \end{equation}
For scale-free profiles, this implies that  $\sigma^2\propto r^{2-\alpha}$ if $\rho \propto r^{-\alpha}$, so that $\sigma^2 \propto r^{0.5}$ for $\alpha=1.5$. Equation (\ref{eq:hydro_sta}) is of course valid only if the velocity dispersion is isotropic but in fact gives the same asymptotic result as more complicated models.

To show this, we can consider the models of \cite{1993MNRAS.265..250D}, which have an inner slope $-\alpha$ that changes at a characteristic radius, $r_{\rm d}$, to an outer slope of $-4$, not too far from what we found in our simulations:
  \begin{equation}
  \rho(r) = \left(r/r_{\rm d}\right)^{-\alpha}\left(1+r/r_{\rm d}\right)^{\alpha-4}
  \end{equation}
  The parameter, $\alpha$, is restricted to the interval of $0 \leq \alpha < 3$. Using the Jeans equation and assuming a constant velocity anisotropy, $\beta=1-\sigma^2_{\rm t} / 2\sigma^2_{\rm r}$, where $\sigma_{\rm t}$ and $\sigma_{\rm r}$ are the tangential and radial velocity dispersions, one finds 
  \begin{eqnarray}
  \sigma_{\rm r}^2 &=& \frac{M_\infty}{r_{\rm d}}\left(\frac{r}{r_{\rm d}}\right)^{-2\beta}\left(\frac{r}{r+r_{\rm d}}\right)^\alpha  \left(\frac{r+r_{\rm d}}{r_{\rm d}}\right)^4 \,\,\nonumber\\
  && \times\quad (-1)^{2\beta}\,{\rm B}\left(-\frac{r_{\rm d}}{r};5-2\beta,2\alpha-6\right),
  \end{eqnarray}
  where $M_\infty$ is the total mass (finite for $\alpha<3$), and ${\rm B}(x;\,a,b)$ is the incomplete Beta-function. It is easy to show that the asymptotic behaviour for $r\ll r_{\rm d}$ in the case $\beta < \alpha-1$ is
  \begin{equation}
  \sigma_{\rm r}^2 \to \frac{M_\infty}{2(\alpha-\beta-1)r_{\rm d}}\left(\frac{r}{r_{\rm d}}\right)^{2-\alpha},
  \end{equation}
which is independent of the anisotropy parameter, $\beta$, and yields $\sigma_r^2 \sim r^{0.5}$ for $\alpha=1.5$. This independence from the velocity anisotropy might contribute that even under perturbations of the velocity anisotropy, the asymptotic velocity dispersion remains unchanged. Clearly, these are crude arguments, and a more in-depth theoretical analysis is necessary, which is however beyond the scope of this paper.
  
The almost purely power law behaviour of these systems reveals another curious property, which we report as a side note. The asymptotic behaviour in the case $\beta>\alpha-1$ can be shown to be $\sigma_r^2\to (M_\infty / r_{\rm d})\,B_0(r/r_{\rm d})^{\alpha-2\beta}$, where $B_0$ is a pre-factor containing Gamma-functions of $\alpha$ and $\beta$. The transition between the two regimes is through an isothermal profile, where $\beta$ fixes $\alpha$. For radially biased orbits with larger $\beta$, the asymptotic velocity dispersion profile thus depends on both $\beta$ and $\alpha$. 
Once $\beta$ has fallen below $\alpha-1$, the velocity dispersion profile then becomes independent of $\beta$ and has the same radial dependence with the gravitational potential, $r^{2-\alpha}$, as can already be seen from solutions to equation~(\ref{eq:hydro_sta}). 
This means that the system shows asymptotically a polytropic behaviour with an effective pressure $P\propto \rho^{2/3}$ (for $\alpha=1.5$; $\gamma=2-2/\alpha$). This has a negative internal energy equal to the negative kinetic energy, which is a result of the virial theorem and the resulting negative specific heat responsible for the gravothermal catastrophe. In contrast to an ideal gas, the pressure thus increases more slowly than the density, leading to a decreasing temperature, or $\sigma^2$, towards the centre. 

\begin{figure}
  \centering 
   \includegraphics[width=85mm]{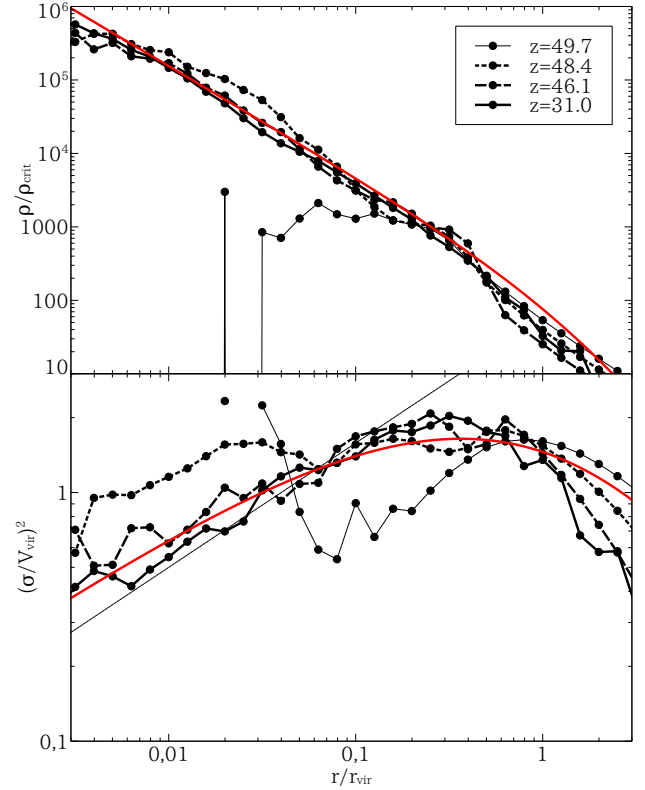}
   \caption{
     Evolution of the radial density (upper) and velocity dispersion (lower) profiles. 
     The initial overdense patch is characterized as $\epsilon=0.05$, $r_{\rm c}/r_{\rm edge}=0.4$ and $g_{\rm amp}=0$ and the collapsed system has the inner density slope of $\alpha=1.5$ (see e.g. Figure \ref{fig:rc_dens}).
     Black thin dashed line in the lower panel represents the expected radial dependence of $\sigma^2 \propto r^{1/2}$. 
     Black thin solid, dotted, dashed and thick solid lines are the snapshots at $z=49.7$, 48.4, 46.1 and 31.0, respectively. 
     Red line represents the Dehnen model of $\alpha=1.5$, assuming the concentration parameter, $c = r_{\rm vir}/r_{\rm d}=3$, which is consistent with cosmological $N$-body simulations with the cut-off \citep{2014ApJ...788...27I}.
     The central part of the system rapidly achieves the equilibrium state of $\rho \propto r^{-1.5}$ and $\sigma^2 \propto r^{1/2}$ after dynamical relaxation ($z \la 46$) and the profiles well match the Dehnen model. 
     \label{fig:evo_dehnen}
}
\end{figure}

Figure \ref{fig:evo_dehnen} presents the evolution of the density and velocity dispersion profiles of the DM halo formed in the run of $\epsilon=0.05$, $r_{\rm c}/r_{\rm edge}=0.4$ and $g_{\rm amp}=0$. Although the condition of the timescale of two-body relaxation is satisfied at $r/r_{\rm vir} \ga 0.02$ at $z=31$, as shown in the previous figures, the profiles are demonstrated over a broader radial range to see how they fit the Dehnen model. We observed the central density structure of $\alpha = 1.5$ in the run (see e.g., Figure \ref{fig:rc_dens}) and the velocity dispersion profile has the expected radial dependence, $\sigma^2 \propto r^{0.5}$ (thin dashed line in the lower panel). The both profiles well fit those of the Dehnen model with $\alpha=1.5$ (red). As illustrated in this figure, after collapse ($z \sim 48$), the central part of the system relaxes and achieves the equilibrium state, rapidly ($z \la 46$), in both the radial density and velocity dispersion profiles.

Another qualitative argument for the persistence of $\alpha=1.5$ profiles can be made based on the rotational support of these haloes. Naively, the rate of change of specific angular momentum vector, ${\bf L}$, i.e. the torque, of the entire system can be written as
\begin{equation}
  \frac{d{\bf L}}{dt} = \int \rho ({\bf r} \times {\bf a}) dV \biggl / \int \rho dV, 
\end{equation}
where ${\bf r}$ and ${\bf a}$ are the position and acceleration vectors and $dV$ represents the volume element, respectively. After the first shell crossing, 
${\bf r} \times {\bf a}$ of each particle is not 0 actually because the symmetry in the initial state is strongly broken. The amount of angular momentum that a particle at $r$ obtains in a time interval, $\Delta t$, is $\Delta L \propto r a \Delta t$. Assuming a power-law density $\rho \propto r^{-\alpha'}$, we obtain the relation, $\Delta L \propto r M(<r)/r^2 \Delta t \sim r^{-\alpha'+2} \Delta t$. Hence, a large fraction of the angular momentum originates at large radii when $\alpha' < 2$, including the central part of DM haloes we study in this paper. 

\begin{figure}
  \centering 
   \includegraphics[width=85mm]{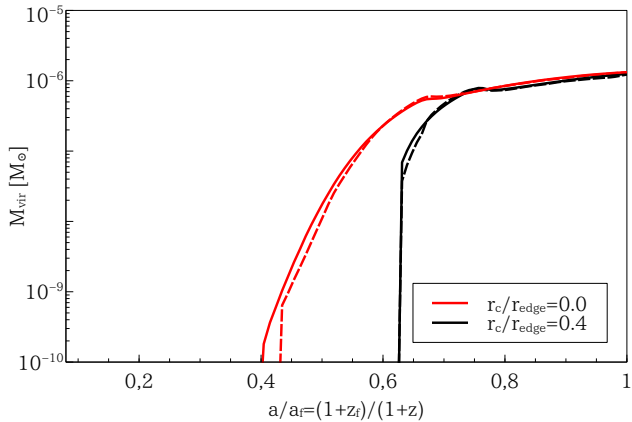}
   \caption{
     Mass accretion history of microhaloes ($\epsilon=0.05$).
     Red and black lines represent the runs of $r_{\rm c}/r_{\rm edge} = 0.0$ and 0.4. Solid and dashed are the runs without ($g_{\rm amp}=0$) and with the noise ($g_{\rm amp}=2$), respectively.  
     In the run of $r_{\rm c}/r_{\rm edge} = 0.4$, the halo formation is delayed but the virial mass at the redshift to finish the simulations, $z_{\rm f}$, is almost same with that in the run of $r_{\rm c}/r_{\rm edge} = 0.0$ because mass accretes in a shorter timescale. 
     The Gaussian noise does not significantly change the mass accretion history as a whole.
     \label{fig:mah}
}
\end{figure}

Figure \ref{fig:mah} illustrates the mass accretion histories of DM haloes. 
The mass contained in the initial core collapses at the same time in the run of $r_{\rm c}/r_{\rm edge} = 0.4$. The halo in the run with $r_{\rm c}/r_{\rm edge} = 0.0$ experiences a more fluent mass accretion history and the material spends a longer time at large radii before collapsing. As a consequence, particles contained in the halo in the run with $r_{\rm c}/r_{\rm edge} = 0.0$ would have higher angular momentum than their counterparts in the run with $r_{\rm c}/r_{\rm edge} = 0.4$. 
Comparing dashed lines with solid ones, the Gaussian noise does not significantly change the mass accretion histories as a whole, although it disturbs the formation of the haloes and alters their central density structure.

\begin{figure}
  \centering 
   \includegraphics[width=85mm]{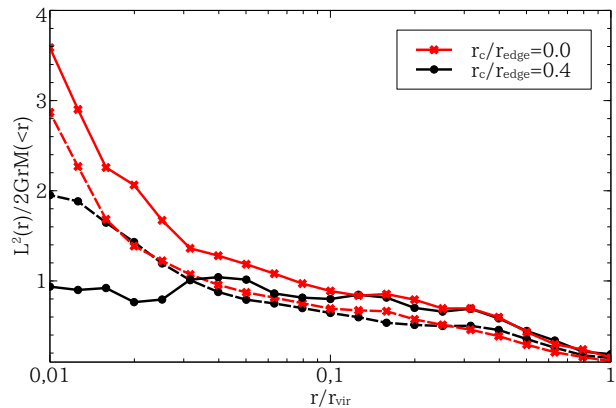}
   \caption{ 
     Degree of rotation support in the runs with $\epsilon = 0.05$. 
     The centrifugal force, $L^2/2r^3$, is compared with gravity, $GM(r)/r^2$.
     Red and black lines are the snapshots at $z=31$ in the runs of $r_{\rm c}/r_{\rm edge}=0.0$ and 0.4.
     Solid and dashed lines represent the results from the runs without ($g_{\rm amp}=0$) and with the noise ($g_{\rm amp}=2$), respectively. 
       When the noise is not imposed, the central part of the collapsed system in the run of $r_{\rm c}/r_{\rm edge}=0.0$ (0.4) is highly (reasonably) rotation supported. 
       The ratio of the centrifugal force to gravity is decreased (increased) in the runs of $r_{\rm c}/r_{\rm edge}=0.0$ (0.4). As a consequence, the degrees of rotation support in the centre are similar in the two runs as with their central density structure. 
     \label{fig:l2_vs_rm}
}
\end{figure}

In Figure \ref{fig:l2_vs_rm}, we compare the centrifugal force with gravity and show that the centrifugal force in the run of $r_{\rm c}/r_{\rm edge}=0.0$ more significantly supports the system than that in the run of $r_{\rm c}/r_{\rm edge}=0.4$ when the Gaussian noise is not included ($g_{\rm amp}=0$; solid lines). In the runs with the noise ($g_{\rm amp}=2$; dashed lines), the ratio of the centrifugal force to gravity is decreased in the runs with $r_{\rm c}/r_{\rm edge}=0.0$, and increased in those with $r_{\rm c}/r_{\rm edge}=0.4$, and the degree of rotational support in the centre is similar in the two runs.   
Although the comparison is not precise actually because the collapsed systems have deviated from the initial (almost) spherical configurations as shown in Figure \ref{fig:image_rc_gamp}, we employ it for simplicity and it helps us to qualitatively understand the difference in the density profile.

Looking at the radial density profile of the haloes in the runs without the Gaussian noise (e.g. Figure \ref{fig:rc_dens}), the density slope gets shallower at the centre in the run with $r_{\rm c}/r_{\rm edge}=0.0$ where the degree of rotational support increases. As argued above, free fall motion of DM particles seems the key to form the density structure of $\alpha = 1.5$ and systems with higher tangential velocity dispersion are expected to have shallower central cusps (e.g. \citealp{2000ApJ...538..517S}; L06; \citealt{2011ApJ...743..127L}). The central part of the halo in the run with $r_{\rm c}/r_{\rm edge}=0.0$ is indeed highly rotationally supported, implying significant tangential motion. Because of the velocity structure, a shallower cusp has been formed in the run of $r_{\rm c}/r_{\rm edge}=0.0$. On the other hand, this toy model works for the run with $r_{\rm c}/r_{\rm edge}=0.4$ in which gravitational collapse happens in a short timescale and angular momentum is lower, implying free fall motion. As expected from these arguments, a density structure of $\alpha = 1.5$ is formed at the centre of the DM halo.

Figure \ref{fig:l2_vs_rm} also shows that in the runs with the Gaussian noise, the degree of rotational support in the haloes is almost the same at $\ga 0.02 r_{\rm vir}$. Although they experience different mass accretion histories (Figure \ref{fig:mah}), their density profiles are similar to each other and the central slope is $\alpha \sim 1$, like that in the NFW model (Figures \ref{fig:bino_gaussian_rc0.40} and \ref{fig:bino_gaussian_rc0.00}). This similarity is clearly caused by the impact of the Gaussian noise which is not taken into account in the toy model and it might in fact be a key to understand the origin of the universal NFW density profile.

\begin{figure}
  \centering 
   \includegraphics[width=85mm]{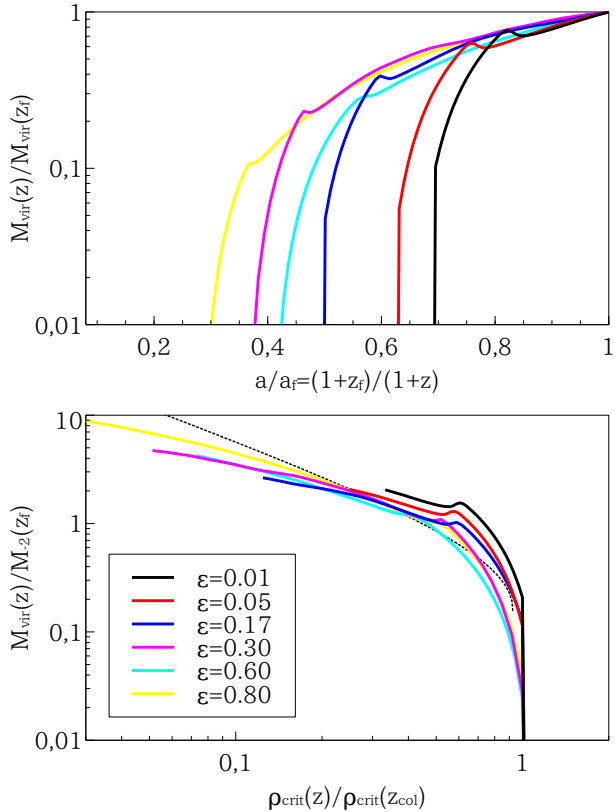}
   \caption{
     Mass accretion history of DM haloes in runs of $r_{\rm c}/r_{\rm edge}=0.4$ and $g_{\rm amp}=0$, varying the parameter to control the initial density slope beyond the core, $\epsilon$. In each panel, black, red, blue, magenta, cyan and yellow lines represent the results of runs of $\epsilon=0.01, 0.05, 0.17, 0.30, 0.60$ and 0.80, respectively. 
     {\it Upper panel:} Mass accretion history is presented as a function of the scale factor, $a$, and the virial mass of the haloes is normalized by that at the final redshift, $z_{\rm f}$ (for details see Table \ref{tab:zi_zf}).  
     {\it Lower panel:} Mass accretion history as a function of the critical density at given redshift, $\rho_{\rm crit}(z)$, scaled by that at the collapse redshift, $z_{\rm col}$, when the virial mass of the halo becomes non-zero. The virial mass of haloes is scaled by the enclosed mass within the radius, at which the density slope is -2, $M_{-2}$, computed at $z_{\rm f}$. Black dashed line represents the Dehnen model of $\alpha=1.5$. 
     \label{fig:mah_eps}
}
\end{figure}

As discussed above, the mass accretion history of DM haloes is an important factor to determine their internal structure. Cosmological $N$-body simulations revealed that the mass accretion history of DM haloes is closely related with their mass profile \citep[e.g.][]{2013MNRAS.432.1103L, 2017MNRAS.465L..84L}. Figure \ref{fig:mah_eps} studies the relation between the mass accretion history and the mass profile of haloes in the runs of $r_{\rm c}/r_{\rm edge}=0.4$ and $g_{\rm amp}=0$, varying the initial density slope beyond the core, $\epsilon$. The central density structure of $\alpha=1.5$ is obtained in these runs (see Figure \ref{fig:lu_diagram}). In the upper panel, the mass accretion history is represented as a function of the scale factor, $a$, normalized by that at the final redshift of the run, $z_{\rm f}$. These haloes experience rapid mass accretion in the early phase because the mass contained in the initial core collapses at the same time and a similarity in the mass growth in the later phase can be found.

Motivated by \cite{2013MNRAS.432.1103L} who found that scaling the mass accretion history of DM haloes in an appropriate way, it reproduces the mass profile of the halo, in the lower panel, we show the mass accretion histories as a function of the critical density at a given redshift, $\rho_{\rm crit}(z)$. The virial mass is scaled by the enclosed mass within the radius at which the logarithmic slope is -2 at $z_{\rm f}$, $M_{-2}(z_{\rm f})$. We scale $\rho_{\rm crit}(z)$ by that measured at the collapse redshift although \cite{2013MNRAS.432.1103L} used that measured at $z_{-2}$ when the virial mass of the halo is equal to $M_{-2}(z_{\rm f})$. The mass accretion histories derived from the simulations resemble each other. The dashed line represents the Dehnen model of $\alpha=1.5$. For this line, the vertical and horizontal axes are the scaled mass enclosed within a radius, $M(<r)$, and mean density, $<\rho> \propto M(<r)/r^3$, respectively. The simulation results may not perfectly fit the Dehnen model although it works well for the density and velocity dispersion profiles. It would be interesting to investigate this relation in more detail in cosmological simulations with the cut-off \citep[see also][]{2016MNRAS.460.1214L}.

\section{Summary and Conclusion}
\label{sec:summary}

The random motion of thermally produced DM particles suppresses the growth of density fluctuations on small scales due to free streaming and thus imposes a cut-off in the matter power spectrum. 
This cut-off scale directly sets the mass scale of the smallest DM haloes that will form. 
Cosmological $N$-body simulations which resolve this cut-off produce central density cusps in DM haloes near the cut-off scale that are steeper, i.e., $\rho \propto r^{-\alpha}$ with $\alpha > 1$, than that of the NFW profile, $\alpha=1$, and generally resemble more a single power law. However, the origin of such steep cusps has remained unclear. The situation is furthermore complicated by the fact that no such power-law cuspy haloes have never been detected in vanilla CDM simulations, where density perturbations are included down to the Nyquist mode of the particle distribution. A resolved cut-off however implies fundamentally different dynamics: haloes that form at the cut-off scale collapse monolithically. This is in stark contrast to the common notion of hierarchical structure formation where all structure, such as galaxies and galaxy clusters, assembles through mergers of smaller scale objects (which would be the case in an ideal, perfectly cold limit). Once such a halo has collapsed from the primordially smooth density field -- by first collapsing to a pancake, a filament, and finally along the third axis -- it will grow further by both smooth accretion as well as mergers with other such primordial haloes. The primordial haloes are thus the atomic building blocks of subsequent hierarchical formation, and an understanding of their density profile and internal dynamics is of importance to understand the development of the systems that develop in later stages.

In this paper, motivated by the spherical infall model developed by the previous analytical studies, we presented a suite of high-resolution, idealised $N$-body simulations of the collapse of proto-halo peaks to haloes in order to elucidate the connection between the properties of the proto-halo patch and the properties of the halo to which it collapses. We focused on the impact of three qualitatively different aspects: its profile, shape and graininess, as approximations to proto-halo patches arising in cosmological Gaussian random fields. 

We found a clear difference in the resulting density profile between proto-halo patches that have a power-law profile (i.e. no cut-off) and those that are initially cored (i.e. with a cut-off):
\begin{itemize}
\item cored proto-haloes {\em universally} lead to power-law like final halo profiles with $\alpha\sim1.5$ independently of the slope of the profile beyond the core and large-scale non-spherical perturbations
\item power-law proto-haloes, in contrast, agree well with the theoretical prediction of L06, i.e., $\alpha = 1.0$ for $\epsilon \la 0.2$, and steeper cusps for $\epsilon\ga0.2$, where $\epsilon$ describes the slope of the proto-halo power law \citep[see][]{1984ApJ...281....1F}
\end{itemize}

Furthermore, we studied the resilience of the central density cusps to the dynamical impacts of small scale perturbations, motivated by the results of recent papers \citep[]{2014ApJ...788...27I, 2016MNRAS.461.3385O, 2016arXiv160403131A}. They demonstrated that mergers can reduce steep cusps of $\alpha=1.5$ and bring it to the {\it universal} NFW-form, $\alpha=1$, that appears more resilient to mergers. Complementing these previous studies, we have focused on much weaker and smaller scale perturbations. 
\begin{itemize}
\item cored proto-haloes are quite sensitive to the impacts of the noise, making the final profile shallower than that of the NFW model in extreme cases.
\item power-law proto-haloes are much more robust to the dynamical impacts of minor mergers and noise, compared with the counterparts in the runs with the cut-off, with their inner slope always being close to the value predicted for an isotropic dispersion by L06
\end{itemize}
We note that spurious haloes or other sources of noise on scales smaller than the physical cut-off scale might cause similar effects as our noise model. This could be a possible source of concern when studying density profiles in WDM and HDM cosmological simulations and disentangling the dynamical impact of subsequent mergers and perturbations due to spurious haloes and other discreteness noise on the profiles.

Finally, we have provided some phenomenological arguments towards explaining our results.
It appears clear from our results that single power-law profile with  $\alpha \sim 1.5$ are naturally formed in gravitational collapse of collisionless fluids. Supported by our simulations, we argue that the inner parts of the haloes
\begin{itemize}
\item arguably are left-overs of a rapid early free-fall phase which already sets a profile $r^{-1.5}$ prior to shell-crossing
\item behave nearly scale free, with power-law density and velocity dispersion, as well as a constant velocity anisotropy
\item appear to be in hydrostatic equilibrium (possibly with polytropic behaviour), and are reasonably well fit by a Dehnen model
\end{itemize}
Furthermore, haloes that collapse from initially cored profiles experience rapid mass accretion histories and the collapsed material cannot be efficiently torqued to alter the density structure. This is in contrast to the uncored proto-haloes that have more gradual mass accretion histories (shell-crossing earlier, but then growing slower). Material thus spends a longer time at larger radii from the centre, where angular momentum is efficiently originated. As a consequence, the density structure of $\alpha = 1.5$ is not formed and the shallower central density slopes are obtained for $\epsilon \ga 0.3$ in these cases.
     
The results shown in this paper provide a qualitative understanding for the density structures of DM haloes which have been formed through monolithic gravitational collapse and have not experienced any mergers.
However, there remain several important open questions. 
For example, the large-scale cosmic density field can tidally torque haloes during their collapse, thus altering their density structure. The impact of such external torques is certainly an aspect that would be interesting to follow up.

In addition, cosmological $N$-body simulations have notoriously suffered from the presence of many spurious haloes due to artificial fragmentation. 
Directly solving the VP system of equations in numerical simulations may be a good way to overcome the difficulty and it would be exciting to see the density structure of DM haloes in those simulations. 
However, it has been impossible to solve the {\em six-dimensional} VP system of equations due to the computational cost and numerical simulations have been restricted to studying systems with some symmetries to reduce the number of dimensions and computational costs \citep[e.g.][]{1971A&A....11..188J, 1971Ap&SS..13..411C, 1976JCoPh..22..330C, 1983PASJ...35..547F, 1987JCoPh..68..202K, 1988JCoPh..79..184Z, 2000MNRAS.311..377H, 2001JCoPh.172..166F, 2011JCoPh.230.6800M, 2017arXiv170101384H}. 
Recent and future developments in increasing of computational power and in new algorithms may help to achieve sufficient resolution to study the structure of individual haloes in numerical simulations that do not rely on particle methods \citep[e.g.][]{2013ApJ...762..116Y, 2016MNRAS.455.1115H, 2017arXiv170208521T}.  

\section*{Acknowledgments}
We thank our referee, Aaron Ludlow, for a careful reading and his helpful and thoughtful comments. 
We also thank Tom Abel, Raul Angulo, Andreas Burkert, Daisuke Nagai, Adi Nusser, Aseem Paranjape and Romain Teyssier for fruitful discussions and comments, and Tomoaki Ishiyama for sharing his valuable simulation results. A part of numerical computations was carried out on HA-PACS at the Center for Computational Sciences at University of Tsukuba. This study was supported by funding from the European Research Council (ERC) under the European Union's Horizon 2020 research and innovation programme (grant agreement No. 679145, project `COSMO-SIMS').

\bibliographystyle{mn2e}
\bibliography{./ref.bib}

\appendix
\section{Numerical convergence and parameter choice}
\label{sec:appendix}

As demonstrated by many previous studies, the results of $N$-body simulations may sensitively depend on the initial conditions and numerical parameter, such as the number of particles and softening length \citep[e.g.][]{1988ApJ...324..288A, 1991ApJ...374..255K, 1997ApJ...479L..79M, 2002MNRAS.332..971B, 2004MNRAS.350..939B, 2009MNRAS.397..775J, 2015MNRAS.450.3724C}.
The sensitiveness has arisen especially in simulations in which the particles are initially cold, collapse under gravity, and then virialize, like in our simulations. The critical aspects are the perfectly cold initial conditions which are somewhat unforgiving to imprecisions when setting them up.
In order to find the setup of the simulations which gives us trustable results, i.e. is insensitive to the initial conditions and is numerically well converged, we perform a series of test simulations in this Appendix. 

\subsection{Purely radial orbits}
\label{subsec:rad_orb}
The early analytical studies \citep[e.g.][]{1972ApJ...176....1G, 1985ApJS...58...39B} restricted their models to particles collapsing on purely radial orbits.
\cite{1984ApJ...281....1F} found the asymptotic central density structure, $\rho \propto r^{-2}$, for $0 < \epsilon \leq 2/3$. 
\cite{2006ApJ...653...43M} confirmed the predicted structure using $N$-body simulations which were three-dimensional but restricted particles to purely radial orbits.

\begin{figure}
  \centering 
   \includegraphics[width=85mm]{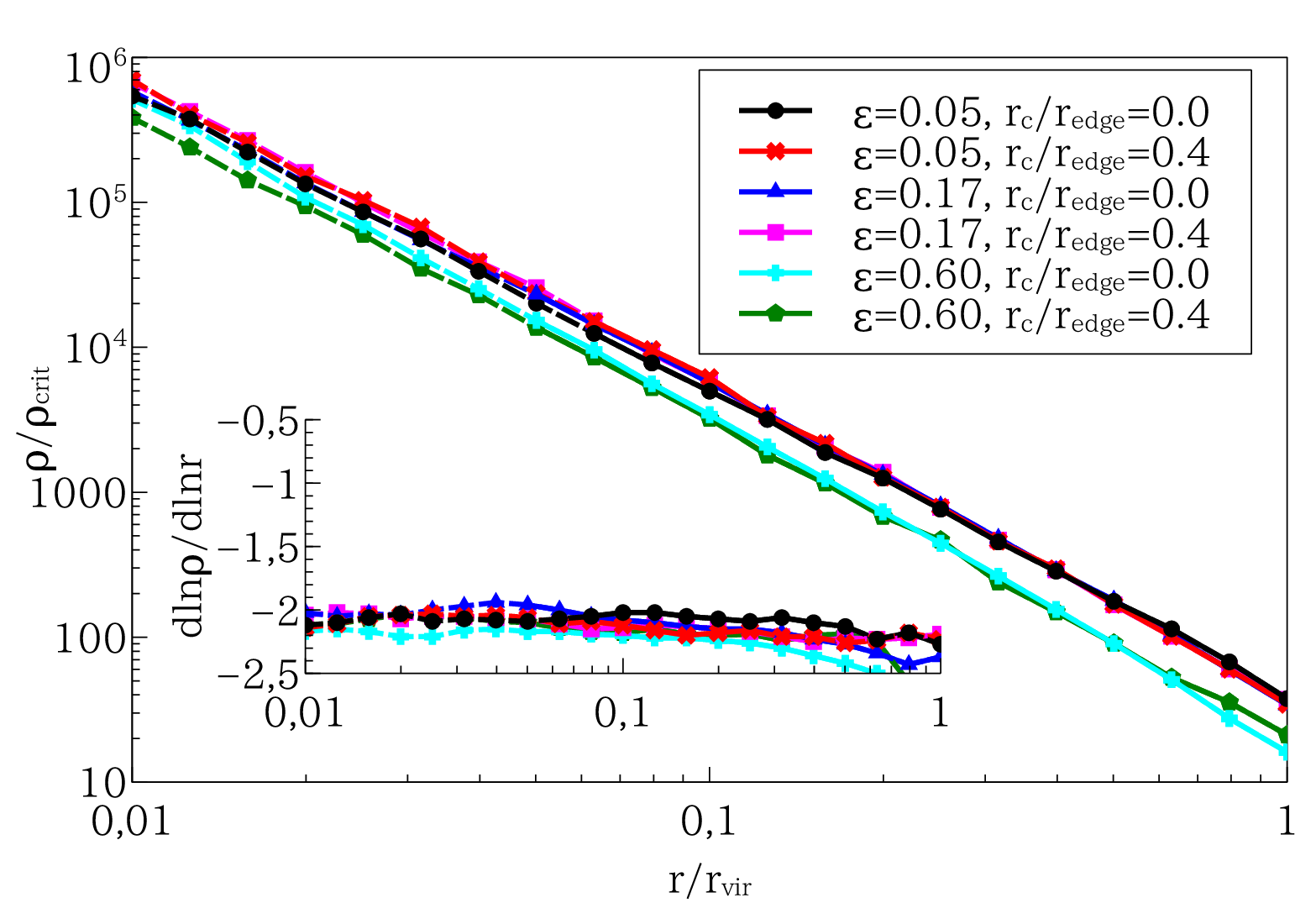}
   \caption{
     Radial density profiles of haloes in the runs with purely radial orbits. 
     The inset panel shows the profile of the logarithmic density slope. 
     The condition, $\tau_{\rm rel}(r) \geq \tau_{\rm sim}$, is (not) satisfied in the radial range of solid (dashed) lines. 
     The resulting profiles are independent of the density structure of the proto-halo patches. 
     As predicted by the anaytical studies, $\rho \propto r^{-2}$ structures are obtained. 
     \label{fig:purely_radial_orbits}
}
\end{figure}

We run the same kind of simulations for a broader range of the initial density structures.
In these runs, as opposed to the runs presented in the main body of the paper, the acceleration vector of each particle, ${\bf a}$, is projected on its position vector, ${\bf r}$.
In other words, the acceleration is given as ${\bf a}_{\rm rad}=-a{\bf r}/r$ and is thus purely radial by definition. 
The initial overdense patches are thus not perturbed by non-spherical terms which arise from particle sampling or numerical approximations. 
We ran simulations varying over a broad range the parameter to control the initial density slope, $0.05 \leq \epsilon \leq 0.60$, for initially cored and non-cored proto-haloes.
Figure \ref{fig:purely_radial_orbits} presents the resulting radial density profiles. Interestingly, almost identical density profiles are obtained in all runs and the density slope is $d\ln{\rho}/d\ln{r} \approx -2$ (see the inset panel), consistent with the analytical prediction.

\subsection{Spherical patches, ROI and convergence}
\label{subsec:spherical}
When we do not perturb the initial density structure with the spherical harmonics with a positive polynomial of degree, $Y_{\rm lm} (l > 0)$, the system initially has a perfect symmetry. 
Perfectly spherical overdense patches should collapse isotropically and form spherical haloes. 
If particles do not have angular momentum initially, results identical to the ones shown in Figure \ref{fig:purely_radial_orbits} would be expected. 

However, it is impossible to obtain such results in three-dimensional simulations with non-radial motions due to two reasons.
The first one is the torque caused by discreteness. 
Numerical simulations have finite resolutions and the actual particle distributions slightly deviate from the spherical symmetry which drives symmetry-breaking through the ROI.
This causes torque which does not exist in spherical systems and makes the orbits of particles non-radial. 
The second reason is due to errors in simulations.  
Any simulation contains errors, due to discretisation of mass and time, and they could also contribute to making the symmetric systems non-symmetric.
The impacts of them would be negligible at the beginning of the simulations, but they can be seeds of radial orbit instabilities and form non-spherical structures. 

\begin{figure*}
  \centering 
   \includegraphics[width=150mm]{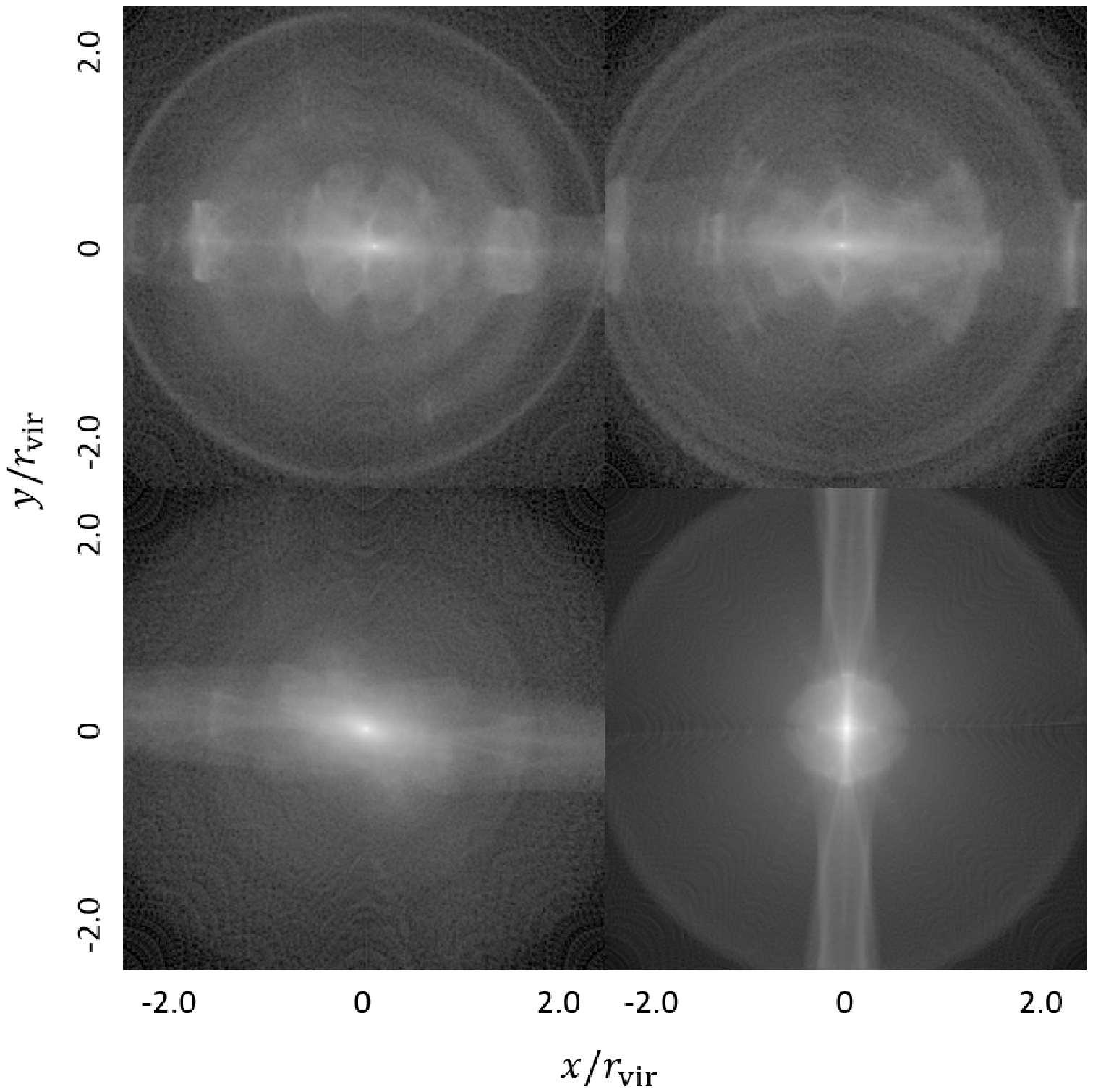}
   \caption{
    Images at $z=31$ in simulations {\em without} the non-spherical perturbation in the initial conditions. 
    The proto-halo patches are described as $\epsilon=0.05$, $r_{\rm c}/r_{\rm edge}=0.4$ and $g_{\rm amp}=0$.
    In the run shown in upper left panel, the reference numerical parameters are adopted, i.e., the softening length, $b=r_{\rm edge}/1280$, opening angle of the tree algorithm, $\theta=0.3$ and number of particles, $N=8,680,336$.
    In the run shown in upper right / lower left / lower right panel, the parameter is varied as $b=r_{\rm edge}/5120$ / $\theta=0.2$ / $N=69,861,728$. 
    Numerically induced and uncontrolled radial orbit instabilities form non-spherical structures. 
    Due to them, the structures in each run are completely different. 
     \label{fig:spherical_image}
}
\end{figure*}

Figure \ref{fig:spherical_image} shows the images obtained in simulations with spherically symmetric initial conditions. 
It is evident that non-spherical structures, which are not expected physically, have been formed through radial orbit instability.
The upper left panel shows the image in the run with the reference numerical parameters. 
In the simulations shown in the other panels, we vary the numerical parameters, the softening length, $b$ (upper right), the opening angle of the tree algorithm, $\theta$ (lower left) and the number of particles, $N$ (lower right). 
It is obvious that they look completely different from each other because the numerically introduced radial orbit instability is not controlled. 

\begin{figure}
  \centering 
   \includegraphics[width=85mm]{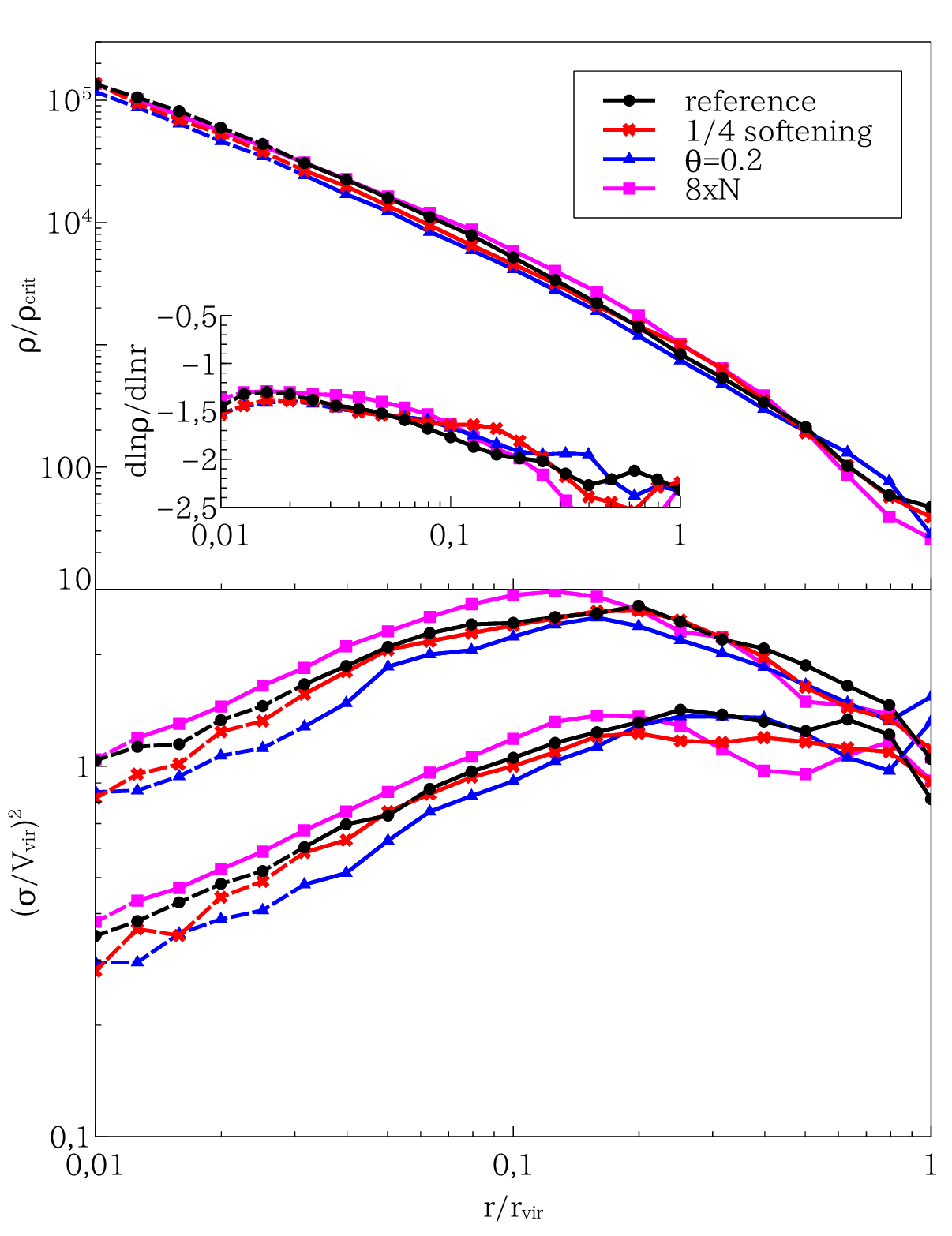}
   \caption{
     Radial profiles of density (upper) and velocity dispersion (lower) in runs {\it without} the non-spherical perturbation, i.e., with a perfect symmetry in the initial conditions.
     The inset panel in the upper one shows the profile of the logarithmic density slope.
     The condition, $\tau_{\rm rel}(r) \geq \tau_{\rm sim}$, is (not) satisfied in the radial range of solid (dashed) lines. 
     In the lower panel, upper (lower) lines show three-dimensional (radial) component, $\sigma^2_{\rm 3d}$ ($\sigma^2_{\rm r}$). 
     The softening length, $b$, opening angle of the tree algorithm, $\theta$, and number of particles, $N$, are varied to test if the results are numerically converged. 
     The profiles are not converged because of numerically introduced, not physical, radial orbit instabilities.
     \label{fig:spherical}
}
\end{figure}

Because of the uncontrolled instability, we cannot know how the results are physically meaningful in these runs. 
The radial profiles of density and velocity dispersion of the collapsed systems are shown in Figure \ref{fig:spherical}.
The density profile (upper) seems to be converged near the centre, but is clearly not in the outskirts of the collapsed systems, where the solution depends strongly on the adopted numerical parameters and the velocity dispersion profile (lower) is not converged at all. 

\subsection{Non-spherical patches}
\label{subsec:non-spherical}
To suppress and avoid the numerical issues described above, non-spherical perturbations are needed that break the symmetry of the problem at a stronger level than that caused by discretisation. Only then it can be hoped that despite the still inherent discretisation of the problem, we can arrive at a converged solution which is then physically meaningful.

\begin{figure*}
  \centering 
   \includegraphics[width=150mm]{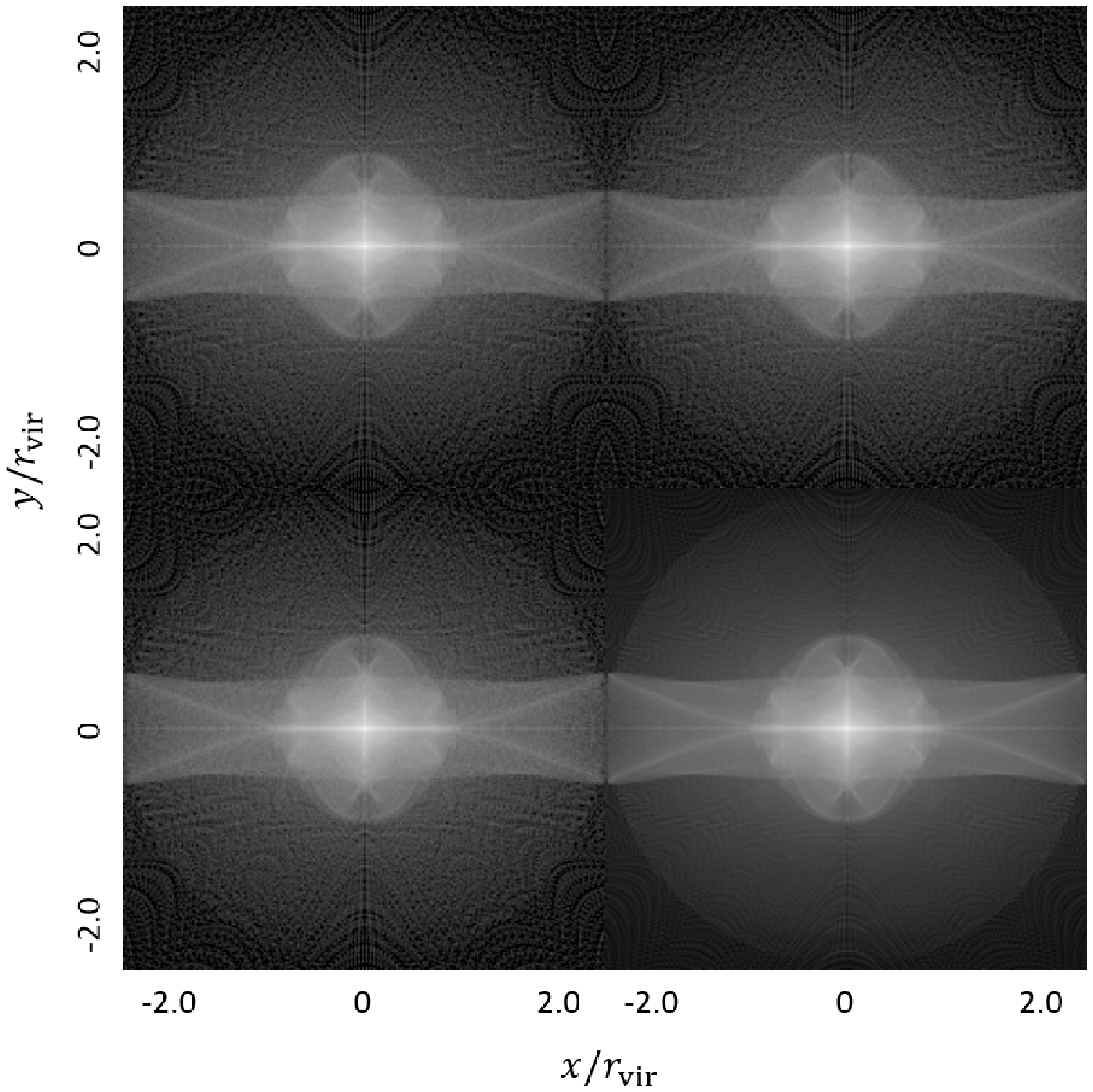}
   \caption{
     Images at $z=31$ in simulations {\em with} the non-spherical perturbation in the initial conditions. 
     The proto-halo patches are described as $\epsilon=0.05$, $r_{\rm c}/r_{\rm edge}=0.4$ and $g_{\rm amp}=0$ and the non-spherical perturbation of $y_{22}=1/4$ is given. 
     In the run shown in upper left panel, the reference numerical parameters are adopted, i.e., the softening length, $b=r_{\rm edge}/1280$, opening angle of the tree algorithm, $\theta=0.3$ and number of particles, $N=8,680,336$.
     In the run shown in upper right / lower left / lower right panel, the parameter is varied as $b=r_{\rm edge}/5120$ / $\theta=0.2$ / $N=69,861,728$.
     The amplitude of the non-spherical perturbation overcomes that of the numerically induced and uncontrolled ones.
     Hence the non-spherical structures are physical in these simulations and are insensitive to varying the numerical parameters. 
     \label{fig:numerical_image}
}
\end{figure*}

In the runs with which we test the numerical convergence, we adopt the same structural parameters as in the runs without the non-spherical perturbation for the initial conditions, $\epsilon=0.05$ and $r_{\rm c}/r_{\rm edge}=0.4$.
The non-spherical perturbation shall be given by the spherical harmonic, $Y_{22}$, and amplitude, $y_{22}=1/4$. 
Figure \ref{fig:numerical_image} shows the images obtained in the simulations with this non-spherical perturbation in the initial conditions. 
In contrast to the ROI dominated results discussed previously, the non-spherical structures that formed can now be considered as physical since they are introduced by the non-spherical perturbation and the features do not change even if we vary the numerical parameters.

\begin{figure}
  \centering 
   \includegraphics[width=85mm]{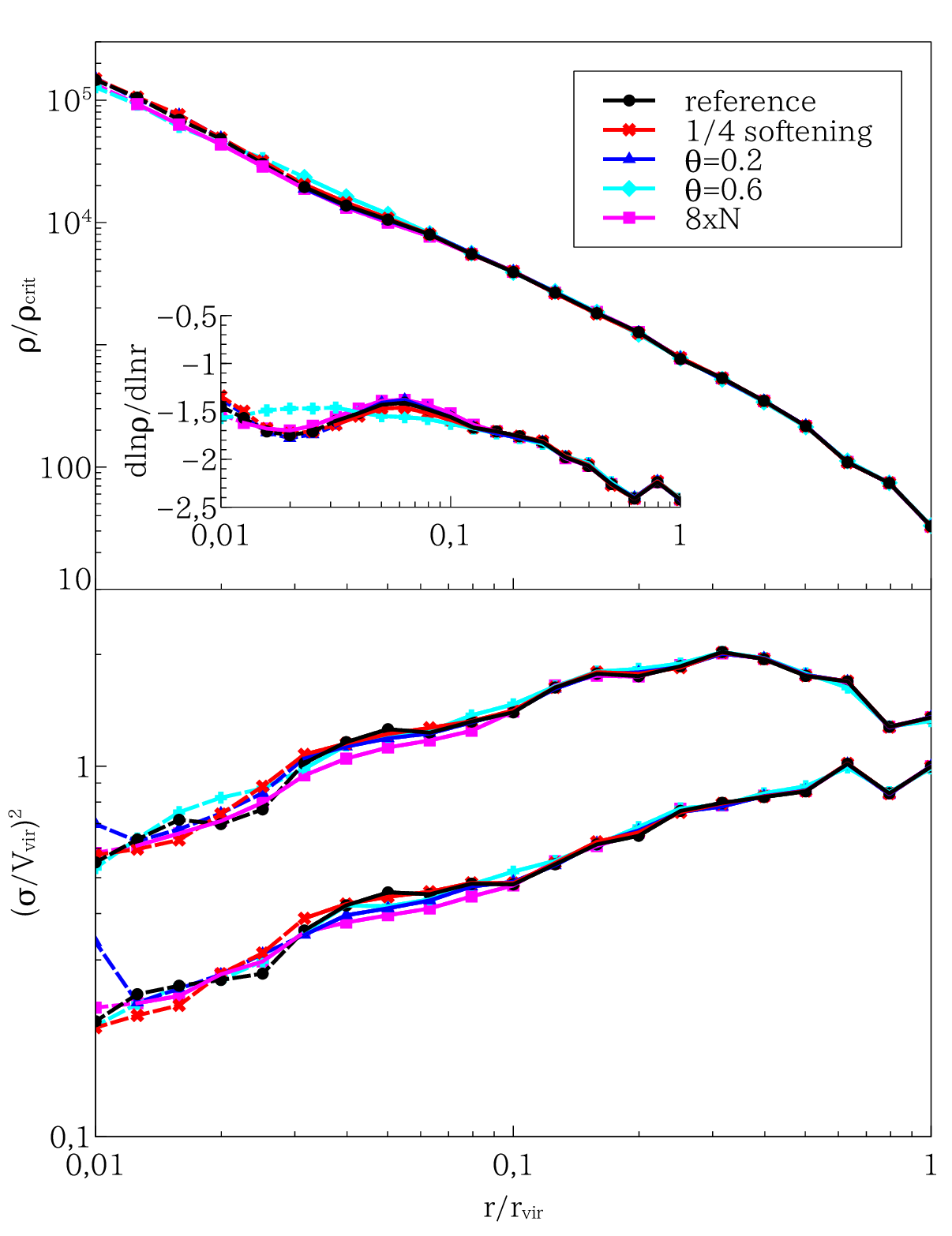}
   \caption{
     Radial profiles of density (upper) and velocity dispersion (lower) in runs {\it with} the non-spherical perturbation.
     The inset panel in the upper one shows the profile of the logarithmic density slope.
     The condition, $\tau_{\rm rel}(r) \geq \tau_{\rm sim}$, is (not) satisfied in the radial range of solid (dashed) lines. 
     In the lower panel, upper (lower) lines show three-dimensional (radial) component, $\sigma^2_{\rm 3d}$ ($\sigma^2_{\rm r}$). 
     The symmetry in the initial condition is broken in some degree by the perturbation. 
     The softening length, $b$, opening angle of the tree algorithm, $\theta$, and number of particles, $N$, are varied to test if the results are numerically converged. 
     Both the density and velocity dispersion profiles are numerically converged when the force is sufficiently accurate ($\theta \leq 0.3$). 
     \label{fig:numerical}
}
\end{figure}

Figure \ref{fig:numerical} depicts the radial profiles of density and velocity dispersion of collapsed systems.
We have achieved much better numerical convergence in the radial profiles with the non-spherical perturbation when the force is calculated with sufficient accuracy (see also Appendix \ref{subsec:param_choice}). 

\subsection{Choice of numerical parameters}
\label{subsec:param_choice}
\begin{figure}
  \centering 
   \includegraphics[width=85mm]{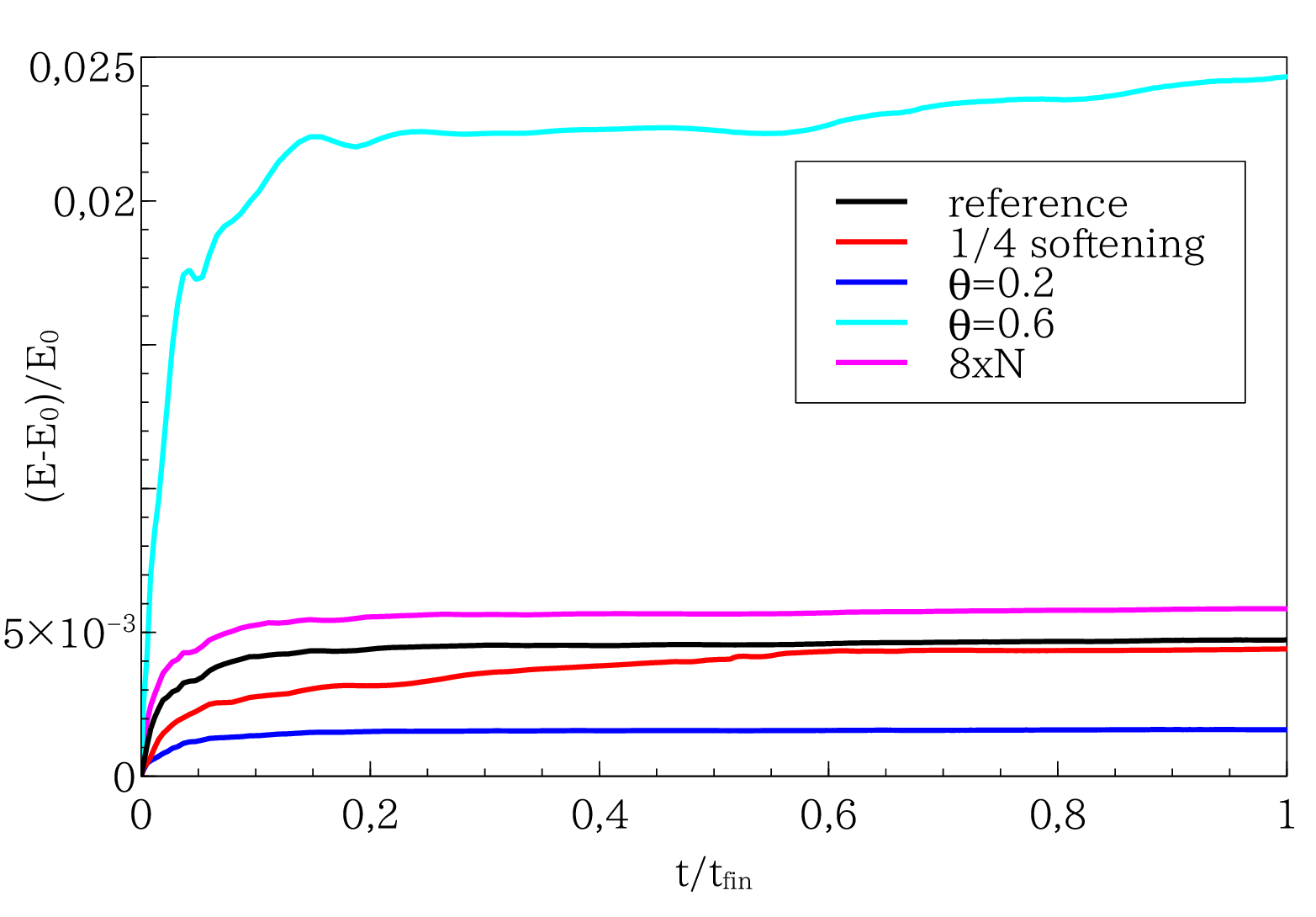}
   \caption{
     Error in the total energy of the $N$-body systems.
     Results of the simulations in Figure \ref{fig:numerical} are shown.
     The dominant fraction of errors in the total energy comes from the force error in the early phase of the simulations at which the overdensity is small and high force accuracy is required to resolve the small inhomogeneousness.
     Hence we need the high force accuracies of $\theta \leq 0.3$ to ensure the suffifient accuracy of the simulation results. 
     \label{fig:numerical_error_in_energy}
}
\end{figure}

Figure \ref{fig:numerical_error_in_energy} shows the error in the total energy of the $N$-body system. 
The results of the same simulations in Figure \ref{fig:numerical} are shown. 
In the reference run (black line), the softening parameter is set as $b = r_{\rm edge}/1280$, and the opening angle $\theta=0.3$ of the tree algorithm is used. 
We use variable time steps following the prescription of \cite{2003MNRAS.338...14P}, $\Delta t \propto \sqrt{b/a_{\rm max}}$, where $a_{\rm max}$ is the maximum absolute value of the acceleration among particles. 
Because of this prescription, $\Delta t$ is reduced to about half its original value over the time of the simulation of the red line. 
One may expect that the error in the total energy gets smaller by a factor of four since we employ a 2nd order time integration scheme, but this actually not the case. 
This is because the dominant part of the error in the total energy comes from the force error during the early phase of the simulations at which the overdensity is small and a high force accuracy is required to resolve the small inhomogeneities. 
Hence the opening angle of the tree algorithm adopted in our simulations, $\theta=0.3$, are smaller than the ones standardly adopted, $\theta=0.6-0.7$ \citep[see e.g.,][]{2003MNRAS.338...14P}. 
Since the radial profiles slightly deviate from each other when we use the standard value, $\theta=0.6$ (cyan), we employ the higher accuracy of $\theta=0.3$. 
In the run of $\theta=0.6$, the error in the total energy exceeds two per cent, although the other runs have errors well below one per cent. 
In addition, the radial profiles and error in the total energy do no change significantly even if the number of particles, $N$, is increased by a factor of eight (from $\sim 9 \times 10^6$ to $\sim 7 \times 10^7$) or the higher force accuracy of $\theta=0.2$ is adopted. 
The chosen numerical parameters thus ensure a sufficient accuracy of the simulation with reasonable numerical costs.

\subsection{Choice of non-spherical perturbations}
\label{subsec:yamp_shape}
\begin{table}
\begin{center}
\caption{
Relation between the amplitude of the non-spherical perturbation, $y_{lm}$, and the axial ratio of the overdense part in the proto-halo patches, $b/a$ and $c/a$, where $a, b$ and $c$ are the three major axes of the overdense region in the patch and satisfy the relation, $a \geq b \geq c$. The axial ratio is derived by solving Poisson's equation. The whole structure of the proto-halo patches is almost spherical (difference in the axial ratio is less than one percent).
}
\begin{tabular}{ccc}
$y_{lm}$ & $b/a$ & $c/a$ \\ 
  \hline
  $y_{22}=1/4$ (reference)             & 0.93  & 0.87 \\
  $y_{22}=1/8$                         & 0.97  & 0.93 \\
  $y_{22}=1/2$                         & 0.87  & 0.76 \\
  $y_{20}=-1/4$ (oblate)               & 1.00  & 0.84 \\
  $y_{20}= 1/4$ (prolate)              & 0.84  & 0.84 \\
\hline 
\end{tabular}
\label{tab:axial_ratio}
\end{center}
\end{table}

\begin{figure}
  \centering 
   \includegraphics[width=85mm]{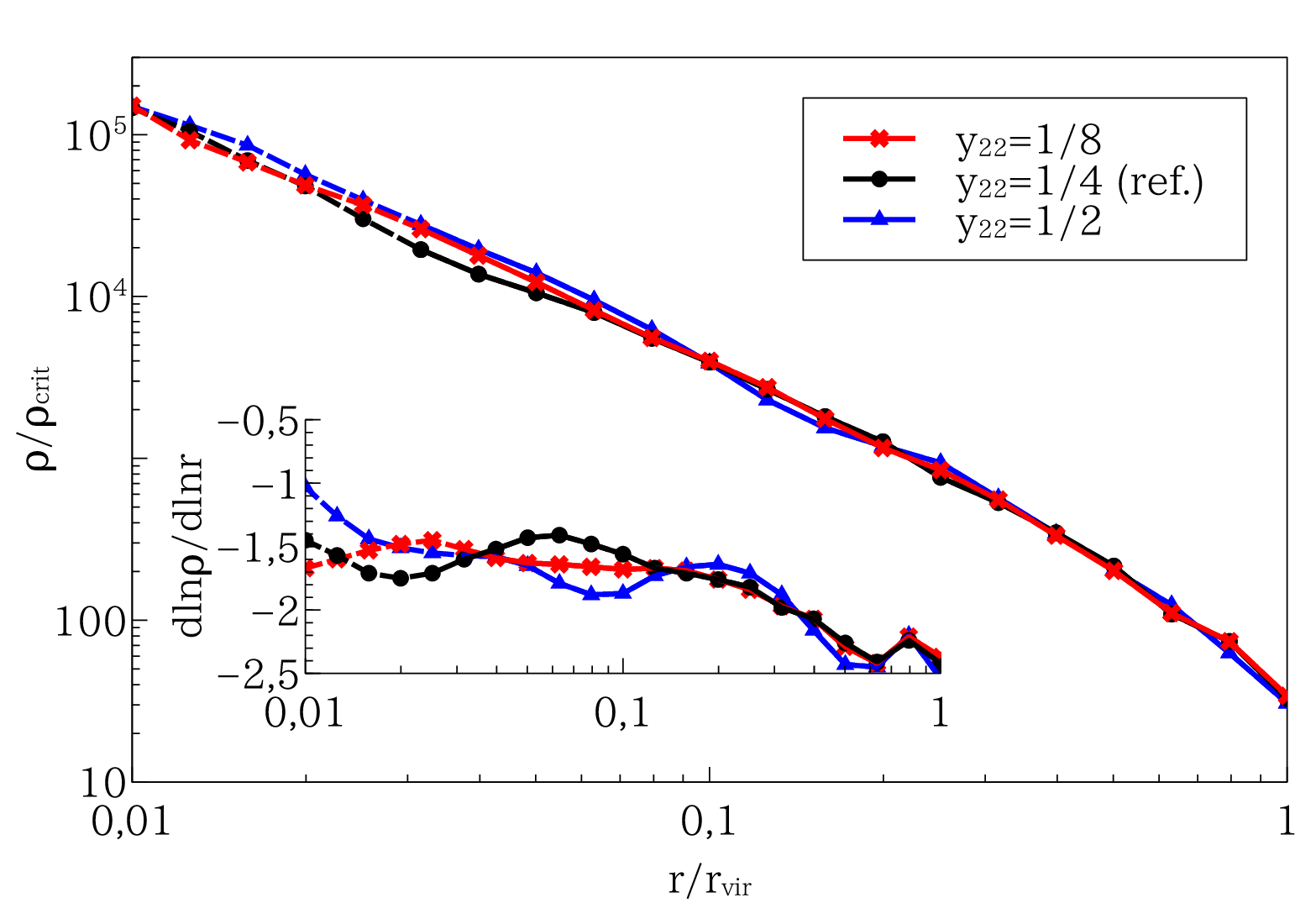}
   \caption{
     Radial density profiles of microhaloes in the runs varying the amplitude of the non-spherical perturbation.
     The inset panel shows the profile of the logarithmic density slope.
     The condition, $\tau_{\rm rel}(r) \geq \tau_{\rm sim}$, is (not) satisfied in the radial range of solid (dashed) lines. 
     We use the perturbation of $Y_{\rm 22}$ in the runs. 
     The density structure is insensitive to the amplitude of the perturbation at $r/r_{\rm vir} \geq 0.02$.
     We adopt $y_{22}=1/4$ as the reference amplitude.
     \label{fig:amplitude}
}
\end{figure}

\begin{figure}
  \centering 
   \includegraphics[width=85mm]{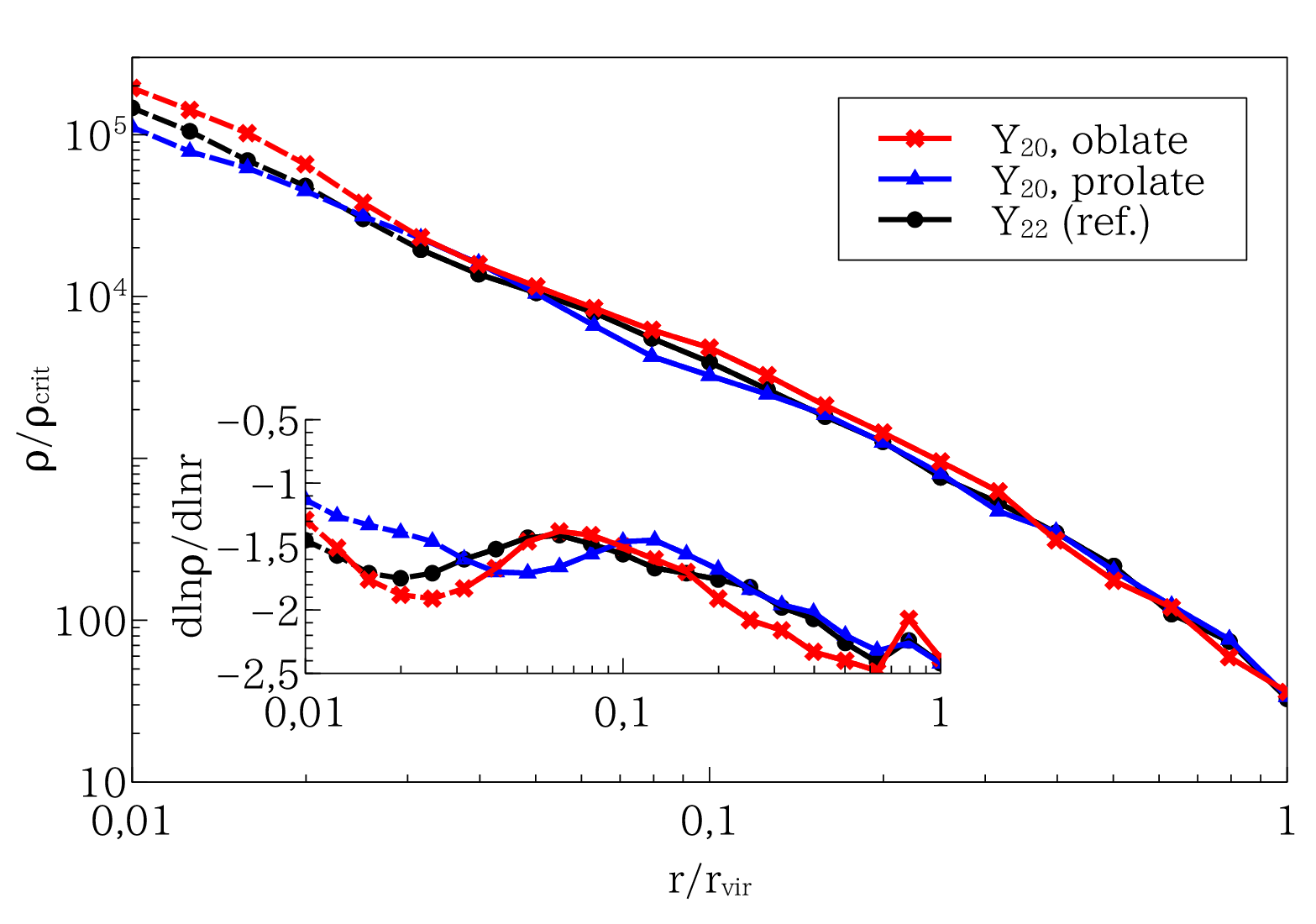}
   \caption{
     Radial density profiles of micorhaloes in the runs varying the shape of the non-spherical perturbation.
     The inset panel shows the profile of the logarithmic density slope.
     The condition, $\tau_{\rm rel}(r) \geq \tau_{\rm sim}$, is (not) satisfied in the radial range of solid (dashed) lines. 
     The amplitude of the perturbation is fixed to be $|y|=1/4$. 
     Runs of red and blue lines adopt the perturbations of the spherical harmonics, $Y_{20}$, with $y_{20}=-1/4$ and 1/4.
     Oblate and prolate perturbations are given in respective models. 
     In the run of black line, the perturbation of $Y_{22}$ with $y_{22}=1/4$ is adopted.
     The density structure is insensitive to the shape of the perturbation at $r/r_{\rm vir} \geq 0.02$.
     We adopt $Y_{22}$-perturbation as the reference model. 
     \label{fig:shape}
}
\end{figure}

By adding the non-spherical perturbation, we introduce additional degrees of freedom in our initial conditions. While the purpose of the perturbation is really only one: to cause a physical symmetry breaking, it may well be that the form and amplitude of the perturbation is reflected in the final solution. If this were the case, then a strong connection between the shape of a proto-halo and its resulting collapsed profile would ensue, which would mean that there is no universal profile to which these proto-halos collapse. So: How strongly do the results depend on the non-spherical perturbation?
To test this dependency, we run two more simulation sets. 
In the first one, the amplitude of the non-spherical perturbation, $g_{\rm amp}$, is varied. 
In the second one, we change the shape of the non-spherical perturbation.
Table \ref{tab:axial_ratio} summarizes the shape of the overdense part in the proto-halo patches and the resulting density profiles are shown in Figures \ref{fig:amplitude} and \ref{fig:shape}. 
It is clear that in these cases, the profiles are insensitive to both the amplitude and the shape of the non-spherical perturbations at $r/r_{\rm vir} \geq 0.02$. This is comforting in the sense that, within the limits we explored, memory of the initial conditions is efficiently wiped out and a universal collapsed profile is obtained. We thus adopt the $Y_{22}$-perturbation with $y_{22}=1/4$ as the reference model for the main part of this paper.

\label{lastpage}
\end{document}